\documentclass[twocolumn,english,prb,notitlepage,superscriptaddress,nobibnotes]{revtex4-2}
\usepackage[T1]{fontenc}
\usepackage[latin9]{inputenc}
\usepackage{graphicx}
\usepackage{amsmath}
\usepackage{float}
\usepackage[colorlinks = true,
linkcolor = blue,
urlcolor  = blue,
citecolor = blue,
anchorcolor = blue]{hyperref}
\usepackage[usenames, dvipsnames]{color}

\usepackage{xr}
\usepackage{braket}
\usepackage{xcolor}

\usepackage{tikz}
\usetikzlibrary{arrows.meta}

\makeatletter
\usepackage{babel}
\usepackage{float}
\usepackage{appendix}

\usepackage{tabularx}
\newcolumntype{C}{>{\centering\arraybackslash}X}

\makeatother

\begin{document}
	
%
%
%
%

\title{Spin wave Hamiltonian and anomalous scattering in NiPS$_3$}

\author{A. Scheie}
\email{scheie@lanl.gov}
\thanks{These authors contributed equally}

\address{MPA-Q, Los Alamos National Laboratory, Los Alamos, NM 87545, USA}

\author{Pyeongjae Park}  
\email{ppj0730@snu.ac.kr}
\thanks{These authors contributed equally}

\address{Center for Quantum Materials, Seoul National University, Seoul 08826, Republic of Korea}
\address{Department of Physics and Astronomy, Seoul National University, Seoul 08826, Republic of Korea}

\author{J. W. Villanova}  
\address{Center for Nanophase Materials Sciences, Oak Ridge National Laboratory, Oak Ridge, TN 37831, USA}

\author{G. E. Granroth}  
\address{Neutron Scattering Division, Oak Ridge National Laboratory, Oak Ridge, Tennessee 37831, USA}

\author{C. L. Sarkis} 
\address{Neutron Scattering Division, Oak Ridge National Laboratory, Oak Ridge, Tennessee 37831, USA}

\author{Hao Zhang} 
\address{Theoretical Division, Los Alamos National Laboratory, Los Alamos, NM 87545, USA}

\author{M. B. Stone}
\address{Neutron Scattering Division, Oak Ridge National Laboratory, Oak Ridge, Tennessee 37831, USA}

\author{Je-Geun Park}
\address{Center for Quantum Materials, Seoul National University, Seoul 08826, Republic of Korea}
\address{Department of Physics and Astronomy, Seoul National University, Seoul 08826, Republic of Korea}
\address{Institute of Applied Physics, Seoul National University, Seoul 08826, Republic of Korea}

\author{S. Okamoto}  
\address{Materials Science and Technology Division, Oak Ridge National Laboratory, Oak Ridge, TN 37831, USA}

\author{T. Berlijn}  
\address{Center for Nanophase Materials Sciences, Oak Ridge National Laboratory, Oak Ridge, TN 37831, USA}

\author{D. A. Tennant}  
\affiliation{Department of Physics and Astronomy, University of Tennessee, Knoxville, TN 37996, USA}
\affiliation{Shull Wollan Center - A Joint Institute for Neutron Sciences, Oak Ridge National Laboratory, TN 37831. USA}

\date{\today}

\begin{abstract}

We report a comprehensive spin wave analysis of the semiconducting honeycomb van der Waal antiferromagnet NiPS$_3$. Using single crystal inelastic neutron scattering, we map out the full Brillouin zone and fit the observed modes to a spin wave model with rigorously defined uncertainty. We find that the third neighbor exchange $J_3$ dominates the Hamiltonian, a feature which we fully account for by \textit{ab-initio} density functional theory calculations. We also quantify the degree to which the three-fold rotation symmetry is broken and account for the $Q=0$ excitations observed in other measurements, yielding a spin exchange model which is consistent across multiple experimental probes. We also identify a strongly reduced static ordered moment and reduced low-energy intensity relative to the linear spin wave calculations, signaling unexplained features in the magnetism which requires going beyond the linear spin wave approximation.

\end{abstract}

\maketitle

\section{Introduction}

\begin{figure}
	\centering
	\includegraphics[width=0.48\textwidth]{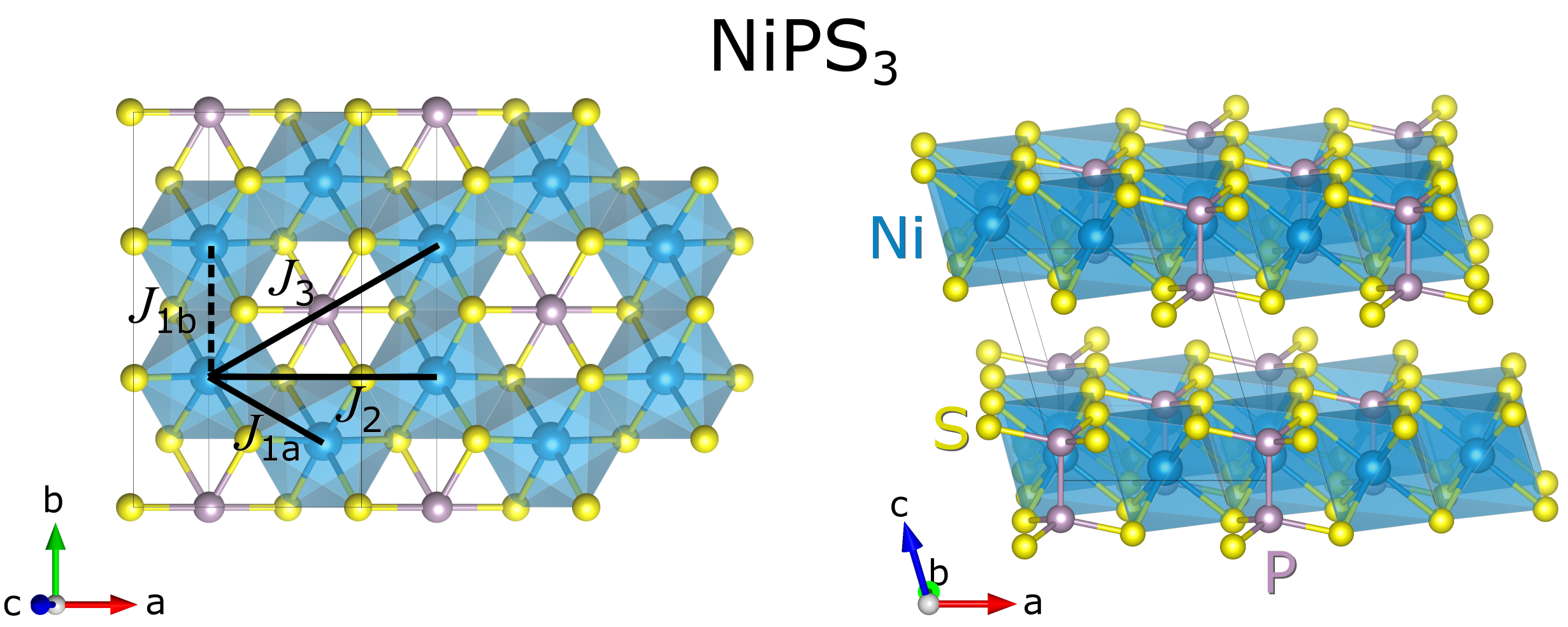}
	\caption{Crystal structure of NiPS$_3$, shown for a single layer (left) and the stacking pattern for multiple layers (right) visualized using \textit{VESTA} \cite{VESTA}. The first three neighbor in-plane exchanges are shown on the left.}
	\label{fig:crystalSchematic}
\end{figure}

Magnetic van der Waals materials which can be exfoliated down to the monolayer limit have tremendous potential for new electronics applications and devices  \cite{Burch2018}. Of special interest is whether new and exotic states can be stabilized because of the low-dimensional properties. 
One such candidate material is NiPS$_3$.  
NiPS$_3$ is a semiconducting layered honeycomb antiferromagnet with the crystal structure shown in Fig. \ref{fig:crystalSchematic}.
Its magnetic Ni$^{2+}$ ions order magnetically at $T_N = 155$~K \cite{Joy_1992,leFlem1982magnetic} to a zig-zag antiferromagnetic order 
with moments along the $c$-axis 
\cite{Wildes_2015}.  NiPS$_3$ has very strong spin-charge coupling \cite{Kim_2018_Charge,afanasiev2021controlling}, and because of this is already being made into workable devices \cite{jenjeti2018field,liu2019nips}. Its magnetic excitations have been measured with powder and single-crystal neutron scattering \cite{Lancon_2018,wildes2022magnetic} and density functional theory shows dominant $J_3$ exchange interaction \cite{Olsen_2021,Mengjuan_2022}, but certain features in its magnetic Hamiltonian (namely the low energy modes) remain imperfectly understood. Perhaps most intriguingly, X-ray, photoluminescence, and optical absorption spectroscopies show a bound exciton state consistent with Zhang-Rice triplet formation between Ni and surrounding S ligands \cite{Kang2020}, which suggests the magnetism is far from conventional. 
This raises two key questions: what is the full exchange Hamiltonian of NiPS$_3$, and are there signs of exotic quantum effects in the collective magnetic excitations?

To answer these questions, we perform a detailed study of single crystal  NiPS$_3$ using inelastic neutron scattering. We fit the spin waves using linear spin wave theory to estimate the magnetic exchange Hamiltonian, perform first principles Wannier function calculations in combination with strong coupling perturbation theory to explain this Hamiltonian, and thus derive a model which accounts for the observed excitations in optical spectroscopy \cite{Kang2020, afanasiev2021controlling}. 
We find a dominant third neighbor exchange (a  behavior which is  unusual but fully explicable with first principles calculations), a strongly reduced ordered moment, and anomalously small low-energy intensity. The third-neighbor exchange is fully explicable with first principles calculations, but the reduced moment and anomalous intensity are not, and thus indicate quantum spin entanglement and higher order effects. 

\section{Experiment and Results}

\begin{figure*}
	\centering
	\includegraphics[width=0.99\textwidth]{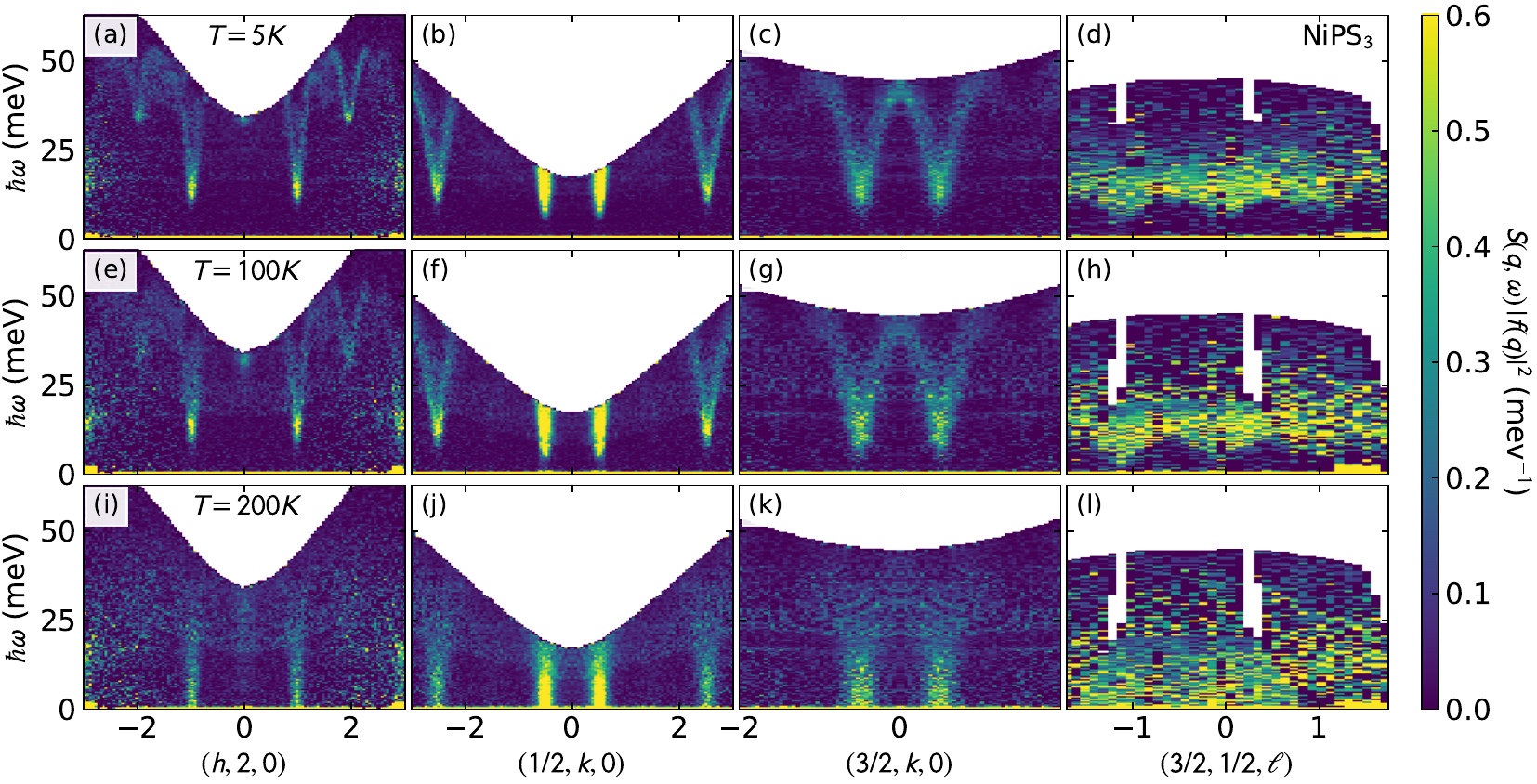}
	\caption{Measured neutron spectra of NiPS$_3$ along different directions in reciprocal space. The top row (a)-(d) shows data at 5~K, the middle row (e)-(h) shows data at 100~K, and the bottom row  (i)-(l) shows data at 200~K. In each panel, the different $E_i$ data are overlaid. The boundaries between the different data sets appear as faint grey lines. At 5~K and 100~K, spin wave modes are clearly visible. The modes become broadened and gapless at 200~K. Note that all data is symmetrized about $h=0$ and $k=0$, and intensity is in absolute units but not corrected for the form factor.}
	\label{fig:inelastic}
\end{figure*}

We measured the inelastic neutron scattering spectrum of NiPS$_3$ using the SEQUOIA spectrometer \cite{Granroth2006,Granroth2010} at the Spallation Neutron Source (Oak Ridge National Laboratory) \cite{mason2006spallation}. The sample consisted of 26 coaligned crystals (total 2.41~g) glued to aluminum plates with the $c$-axis vertical (see Appendix \ref{app:ExpDetails} for details). Although NiPS$_3$ technically has broken three-fold rotation symmetry at the Ni$^{2+}$ sites, the distortion is so weak that we could not easily distinguish $(h00)$ from $(hh0)$, and in the coalignment we treated them as identical. In the plots in this paper, the cuts listed (e.g. in Fig. \ref{fig:inelastic}) in reality include a superposition of cuts rotated in the plane by $\pm 120^{\circ}$. (In the spin wave modeling below, we calculated the three overlapping orientations with a weighting 1:1:1.) For a background, we measured an identical sample holder with no sample. 
We measured the inelastic spectra in the $(hk0)$ scattering plane with incident energies ($E_i$) of 28~meV, 60~meV, and 100~meV at 5~K, and $E_i=28$~meV and $E_i=100$~meV at 100~K, and 200~K. Data were then symmetrized by in-plane reflections about $(h0\ell)$ and $(0k\ell)$. Two-dimensional slices of scattering data are shown in Fig. \ref{fig:inelastic}. 

\begin{figure}
	\centering
	\includegraphics[width=0.48\textwidth]{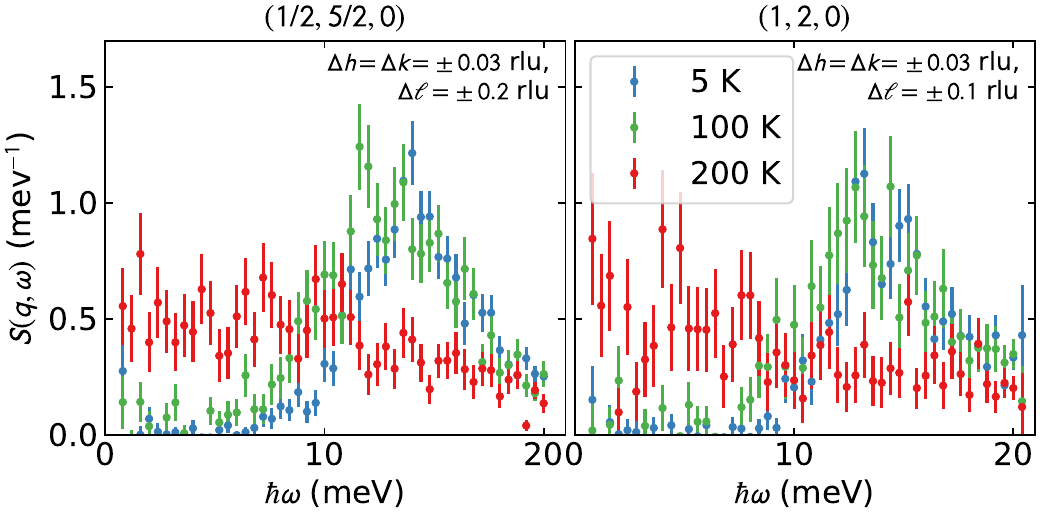}
	\caption{Temperature evolution of the low energy gapped mode in NiPS$_3$, showing $(1/2,5/2,0)$ and $(1,2,0)$ wavevectors measured with $E_i=28$~meV neutrons. Between 5~K and 100~K, the intensity maximum shifts slightly lower in energy, while at 200~K (above $T_N$) the modes become gapless. Note that the intensity profiles of the two points (which nominally correspond to $C$ and $\Gamma$) are identical, even with different $\ell$ integration widths.}
	\label{fig:gapEvolution}
\end{figure}

In the 5~K and 100~K data, spin wave modes are clearly visible in the data, being very well-defined in the in-plane scattering directions, with a pronounced maximum intensity at $\sim 14$~meV. At 200~K (above $T_N = 155$~K), the modes are less well defined and the gap closes. 
The most intense inelastic scattering is at the bottom of the dispersion at $k$ and $h$ wavevectors associated with the zig-zag antiferromagnetic order. This mode has very steeply dispersing magnon modes which, because of experimental resolution broadening, makes the low energy extent difficult to experimentally determine. Nevertheless, as temperature increases the gap steadily closes (Fig.  \ref{fig:gapEvolution}). 
This temperature-dependent gap is well-understood for low-dimensional magnets \cite{nagata1974antiferromagnetic}, and was also observed in FePS$_3$ \cite{Wildes_2012} and MnPS$_3$ \cite{WILDES20071221}.

Although experiments clearly show NiPS$_3$ to be dominated by in-plane exchange interactions, a weak dispersion is visible in the $\ell$ (out-of-plane) direction as shown in Fig. \ref{fig:inelastic}(d). Because of the intense, highly dispersive scattering, the $\ell$ dependence appears as a lower envelope to the scattering  with a bandwidth of $6.5$~meV. The $\ell$ periodicity is the same as the lattice, indicating ferromagnetic inter-planar exchange. 

\section{Spin wave fits}

Having observed such well-defined magnons, we fitted a linear spin wave theory (LSWT) model to the data to determine the exchange constants. 
However, we must also ensure that our model is consistent with other experiments. 
From other studies, it is clear that the Ni$^{2+}$ magnetism is predominantly easy-plane  \cite{Joy_1992,kim2019suppression,Olsen_2021,Mehlawat_2022}. In addition, multiple measurements have reported three low-energy $Q=0$ magnetic modes in the NiPS$_3$ ordered phase: ESR indicates $\Delta_1 = 1.07$~meV \cite{Mehlawat_2022}, optical spectroscopy indicates $\Delta_1 = 1.16$~meV and $\Delta_2 = 3.79$~meV \cite{afanasiev2021controlling}, and photoluminescence indicates $\Delta_1 = 1.7$~meV and $\Delta_2 = 3.3$~meV, deduced from shoulder peaks near the main photoluminescence peak (proposed to be the Zhang-Rice singlet to Zhang-Rice triplet transition) \cite{Kang2020}.
Meanwhile, THz optical spectroscopy reveals a clear $Q=0$ magnon mode at $\Delta_3 = 5.5$~meV which disappears as $T \rightarrow T_N$, and is also present in the absence of photo-excitation \cite{Belvin2021}. By semiclassical spin wave theory, there can only be two low energy $Q=0$ modes in NiPS$_3$. We believe the true $Q=0$ magnons are $\Delta_1$ and $\Delta_3$ because the $\Delta_2 \approx 3.5$~meV mode only appears with $> 1~eV$ optical pumping and is near $2\Delta_1$, suggesting it is a nonequilibrium effect and/or involves the creation of two low energy magnons. (One other possibility is a longitudinal magnon mode, but $SU(3)$ simulations do not find an additional mode, see Appendix \ref{app:Su3}. Furthermore, longitudinal modes are typically far broader in energy than the transverse magnons \cite{Affleck_1989}, not sharp modes as observed in NiPS$_3$.)
Thus, we take the low energy $Q=0$ magnon gaps to be $\Delta = (1.3 \pm 0.3)$~meV (average of the reported values) and $\Delta = (5.5 \pm 0.3)$~meV (uncertainty taken from the lower energy mode). 
We include these fixed average gap values in our model for the low energy $Q=0$ modes. 

The fitted model is based on Heisenberg (isotropic) exchange with single ion anisotropy terms
\begin{equation}
    \mathcal{H} = \sum_{ij} J_{ij} {\bf S}_i \cdot {\bf S}_j + \sum_i \big[A_x ({S^x_i})^2 + A_z ({S^z_i})^2 \big]
\end{equation}
where $\bf{S}_i$ are quantum spin operators of length 1, $J_{ij}$ determine the exchange interaction strengths, and $A_x$ and $A_z$ are the single ion anisotropy terms. 
To constrain the fit, we extracted the spin wave dispersions from the neutron data by fitting multiple independent constant-$Q$ cuts with a Gaussian across 18 different $\hbar \omega$ vs $Q$ slices (see supplemental information \cite{SuppMat}). Where the dispersion was steep, we also fitted constant $\hbar \omega$ slices, 
yielding a total 267 individual $Q$ and $\hbar \omega$ points (treating data from each different measured $E_i$ separately). We then fit the NiPS$_3$ spin waves to the mode energies using \textit{SpinW} \cite{SpinW} assuming three in-plane exchanges and one out-of-plane exchange $J_4$. We found that, in order to produce a $q=K=(1/2,1/2,0)$ intensity maximum at $\sim 14$~meV whilst retaining the two low energy $Q=0$ modes above, the three-fold rotation axis must be weakly broken (which, as shown in Fig. \ref{fig:crystalSchematic}, is true of $C2/m$ space group for NiPS$_3$). 
Otherwise, the only modes at $q=K$ would be $\Delta = 1.3$~meV  and $\Delta = 5.5$~meV. 
We therefore allow the two symmetry-inequivalent first neighbor bonds to have different values ($J_{1a}$ and $J_{1b}$), while the other in-plane exchanges $J_2$ and $J_3$ are assumed to have the same exchange on all honeycomb bonds. 
The resulting fitted parameters are in Table \ref{tab:FittedJ}, with the linear spin wave simulated scattering in Fig. \ref{fig:spinWaveFit}. A plot of higy-symmetry cuts is shown in Fig. \ref{fig:Spaghetti}. (see Appendix \ref{app:SpinWaveFits} for the uncertainty estimation method, and note that although $J_{1a}$ and $J_{1b}$ individual uncorrelated uncertainties overlap, $J_{1b}/J_{1a} > 1$ to within uncertainty.)

One thing that was immediately apparent was that---even assuming a broken three-fold rotation symmetry---there was far too much intensity at low energies for all our initial LSWT simulations [Fig. \ref{fig:spinWaveFit}(f)-(j)]. However, we found that if we calculated the LSWT over a finite window in $Q$ transverse to match the experimental bin widths ($\pm 0.05$ reciprocal lattice units [RLU] in the plane, and $\pm 0.25$ RLU out of the plane), we reproduced the low energy modes better [Fig. \ref{fig:spinWaveFit}(k)-(o)], though not perfectly as we discuss below. The dispersion is so steep there that any finite bin size broadens the modes and shifts the intensity maximum to higher energy transfers. This explains the anomalous intensity down to very low energies at the antiferromagnetic wavevectors also observed in Ref. \cite{wildes2022magnetic}. This also meant that the fitted experimental $Q$ and $\hbar \omega$ points, because they were extracted from cuts with finite bin size, are higher than the actual modes. We therefore calculated the difference in mode energy between the raw LSWT calculation [Fig. \ref{fig:spinWaveFit}(f)-(j)] and the finite-bin summed LSWT calculation [Fig. \ref{fig:spinWaveFit}(k)-(o)], corrected the experimental $Q$ and $\hbar \omega$ points by this offset, and refit the Hamiltonian. The values in Table \ref{tab:FittedJ} represent the fit to these corrected dispersions.

\begin{figure*}
	\centering
	\includegraphics[width=0.99\textwidth]{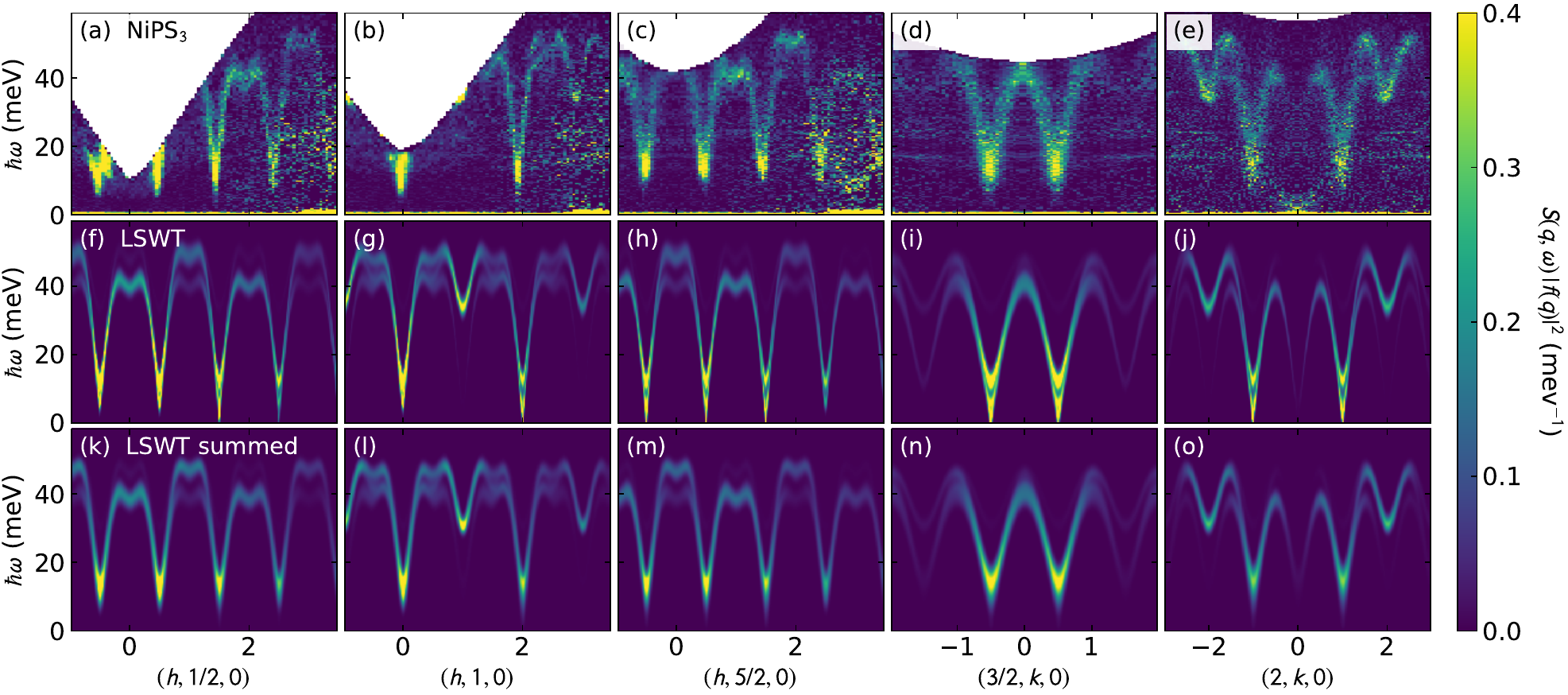}
	\caption{Measured and fitted NiPS$_3$ spin wave spectra. The top row (a)-(e) shows the measured spin wave spectra 
 (with $E_i=100$~meV, $60$~meV, and $28$~meV overlaid as in Fig. \ref{fig:inelastic}).
 The second row (f)-(j) shows the fitted linear spin wave theory (LSWT) spectra, and the bottom row shows same LSWT spectra integrated over the finite widths in $h$, $k$ and $\ell$ for the actual experimental data. Because of finite bin widths, the low energy scattering is much weaker than it would be with infinitesimal bins. (Note that the low-energy intensity in panel (e) near (200) is an acoustic phonon mode.)} 
	\label{fig:spinWaveFit}
\end{figure*}

In fitting the Hamiltonian, we also included in-plane exchange terms beyond the third neighbor in-plane, but we found that these did not improve the reduced $\chi^2$ by $\geq 1$, and thus we consider them to be zero to within uncertainty. We also tried including a Kitaev term in the exchange, but this also did not improve the fit, and instead introduced extra modes in the spectrum which are not present in experiment. Finally, we note that 
in reality, the broken rotation symmetry will affect all bonds, not just $J_1$. However, to reduce the number of fitted parameters we collect all such effects in $J_1$ in order to provide a minimal model for reproducing the experimental observations. 

\begin{table}

\caption{Hamiltonian exchange parameters for NiPS$_3$. The left column shows the best fit model in units of meV, where the broken three-fold symmetry is represented by $J_{1a}$ (two nearest-neighbor exchanges with components along the $a$ axis) and $J_{1b}$ (nearest-neighbor exchange along the $b$ axis). Error bars indicate one standard deviation uncertainty. The right column shows the DFT calculated exchange constants for $J_1$, $J_2$, and $J_3$, which are very close to the experimental values.}
	\begin{ruledtabular}
		\begin{tabular}{c|c|c}
model & fitted LSWT & DFT + perturbation\\
  &   & ($U=3$~eV, $J_H=0.5$~eV) \\
\hline
$A_{x}$  &  $ -0.010 \pm 0.005$ &  \\
$A_{z}$  &  $ 0.21 \pm 0.03$ &  \\
$J_{1a}$    &  $ -2.7 \pm 0.4$   &   -2.7 \\
$J_{1b}$    &  $ -2.0 \pm 0.4$   &   -2.4 \\
$J_2$    &  $ 0.2 \pm 0.3$    &  -0.42 \\
$J_3$    &  $ 13.9 \pm 0.4$   &  13.9 \\
$J_4$    &  $ -0.38 \pm 0.05$    &  \\
	\end{tabular}\end{ruledtabular}
	\label{tab:FittedJ}
\end{table}




One of the most striking features of the fit is that the third neighbor exchange $J_3$ dominates the Hamiltonian. A dominant third-neighbor in-plane exchange is not unusual for hexagonal $3d$ magnets, as seen in e.g. NiGa$_2$S$_4$ \cite{Takubo_2007}, Ba$_2$NiTeO$_6$ \cite{Asai_2017}, Na$_2$Co$_2$TeO$_6$ \cite{yao2022excitations}, BaNi$_2($AsO$_4)_2$ \cite{Gao_2021_Spin}, and BaCo$_2$(AsO$_4$)$_2$ \cite{halloran2022geometrical} and many members of the $M$P$X_3$ family \cite{Chittari2016,Sugita_2018,Kim_2020_Spin}. However, the extremely large $J_3$ we derive (nearly 6 times larger than $J_1$, or $|J_3/\bar{J_1}| = 5.9$) is, to our knowledge, the largest observed.

\begin{figure}
	\centering
	\includegraphics[width=0.47\textwidth]{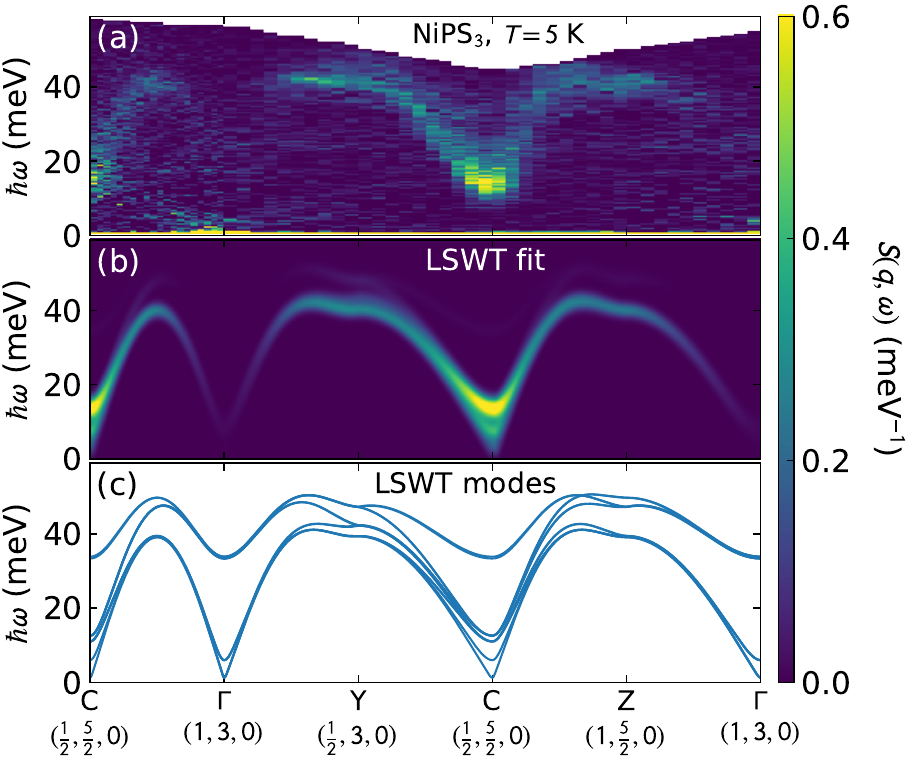}
	\caption{
Plot of scattering data along high-symmetry directions (a) compared to the linear spin wave theory (LSWT) simulation (b) and the LSWT mode energies (c).}
	\label{fig:Spaghetti}
\end{figure}

\section{First principles calculations}

\subsection{Density functional theory}\label{sec:DFT}

To explain this enormous $J_3$, we perform density functional theory (DFT) calculations to estimate the Hamiltonian from first principles. Ref.~\cite{Sugita_2018} studied Dirac cones formed by the half-filled $e_g$ bands in monolayer MBX$_3$, focusing on monolayer PdPS$_3$. They plotted the monolayer Wannier functions and reported the hopping integrals to neighboring transition metal sites. Likewise, we justify the hierarchy of magnetic exchange constant magnitudes by examining the respective hopping integrals of the $e_g$ bands first. The maximum magnitude of the nearest, second, and third-neighbor hopping integrals are 52.94, 29.54, and 215.52 meV, respectively (see also Table~\ref{tab:hopping1N}). The $e_g$-$e_g$ hopping integrals for nearest neighbors are relatively small, and they are even surpassed by the $e_g$-$t_{2g}$ hopping integrals (52.94 $<$ 176.01 meV). The present calculation for bulk NiPS$_3$ is distinct from the monolayer calculation of Ref.~\cite{Sugita_2018} in that it incorporates the $t_{2g}$ orbitals, and their relative importance for the nearest-neighbor exchange is already evident. Next we explain why the inclusion of the $t_{2g}$ orbitals is necessary to capture FM exchange for the nearest neighbors (failure to do so inaccurately leaves one with an AFM $J_1$ and overestimates $J_3$) and examine how the large third-neighbor hopping integral comes to be.

\begin{figure*}
\includegraphics[width=0.9\textwidth]{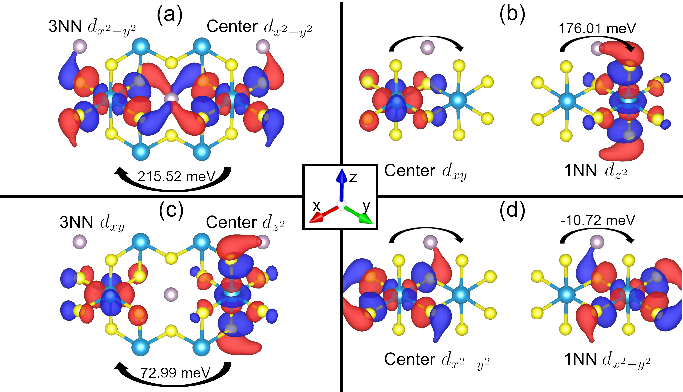}
\caption{Wannier function overlaps. (a) The largest hopping integral for the third-neighbor (3NN) between $d_{x^2-y^2}$ orbitals. (b) The largest hopping integral for the nearest-neighbor (1NN) between $d_{xy}$ and $d_{z^2}$ orbitals. (c) The second largest hopping integral between 3NN $d_{xy}$ and $d_{z^2}$ orbitals. (d) The diminished hopping integral between 1NN $d_{x^2-y^2}$ orbitals. Red (blue) surfaces are the positive (negative) isosurfaces. Teal atoms are Ni, grey atoms are P, and yellow atoms are S (like Fig. \ref{fig:crystalSchematic}).}
\label{fig:overlaps}
\end{figure*}

We visually demonstrate how the hopping integrals are either notable or diminished in Fig.~\ref{fig:overlaps}. The third-neighbor (3NN) $e_g$-$e_g$ hopping integrals are the largest of all, leading to the large AFM 3NN exchange. Previous work has argued for substantial overlap to produce the $d$-$p$-$p$-$d$ exchange for the $e_g$ orbitals for 3NN hopping \cite{Sugita_2018,Sivadas2015} and this is shown in Fig.~\ref{fig:overlaps}(a). The 3NNs do not share ligand S-atoms, and the $p$-tail lobes point toward each other further enhancing overlap. By contrast, the nearest-neighbor (1NN) $e_g$-$e_g$ hopping integrals are relatively diminished, leading to small FM 1NN exchange. As shown in Fig.~\ref{fig:overlaps}(d), the orientation of the $p$-tails on the shared ligand S-atoms is nearly orthogonal leading to substantial cancellation.

The $e_g$-$t_{2g}$ hopping integrals are important both to capture the FM nature of the 1NN exchange and to accurately calculate the 3NN exchange. Fig.~\ref{fig:overlaps}(b) shows the 1NN overlap between $d_{xy}$ and $d_{z^2}$ orbitals. Again, the 1NN share their ligand S-atoms, but the $p$-tails overlap each other to reinforce the hopping. In Fig.~\ref{fig:overlaps}(c), the 3NN overlap between $d_{xy}$ and $d_{z^2}$ orbitals is relatively diminished; the $d_{z^2}$ orbital's primary $p$-tails are not pointing toward the neighboring Ni atom, and the smaller $p$-tails point toward each other less directly than in the $e_g$-$e_g$ case. This $e_g$-$t_{2g}$ process contributes to FM 3NN exchange, but it is merely a small fraction of the large AFM exchange supported by the $e_g$-$e_g$ hopping. 

\subsection{Perturbation Theory}\label{sec:PT}
Armed with the full $e_g$ and $t_{2g}$ tight-binding Hamiltonian, we apply perturbation theory in the strong-coupling limit (explained in Appendix \ref{app:DFTmethods}) to extract the exchange constants, listed in Table \ref{tab:FittedJ}. These are in close agreement with the experimentally fit exchange constants, except for the theoretical prediction that $J_2$ be weakly ferromagnetic where it is experimentally shown to be weakly antiferromagnetic.  Our perturbation theory results are also in close agreement with the results in Ref. \cite{Mengjuan_2022} obtained from fitting total energies of magnetic configurations simulated with DFT+U.

\begin{figure}
\begin{tikzpicture}
	
	\draw (0,0) rectangle (1,1); 
	\draw[->,thick,-{Latex[round,length=2mm,width=2mm]}] (0.33,0.2) -- (0.33,0.8); \draw[->,thick,-{Latex[round,length=2mm,width=2mm]}] (0.67,0.8) -- (0.67,0.2);
	\draw (0,1) rectangle (1,2); 
	\draw[->,thick,-{Latex[round,length=2mm,width=2mm]}] (0.33,1.2) -- (0.33,1.8); \draw[->,thick,-{Latex[round,length=2mm,width=2mm]}] (0.67,1.8) -- (0.67,1.2);
	\draw (0,2) rectangle (1,3);
	\draw[->,thick,-{Latex[round,length=2mm,width=2mm]}] (0.33,2.2) -- (0.33,2.8); \draw[->,thick,-{Latex[round,length=2mm,width=2mm]}] (0.67,2.8) -- (0.67,2.2);
	\draw (0,3.5) rectangle (1,4.5);
	\draw[->,thick,-{Latex[round,length=2mm,width=2mm]}] (0.33,3.7) -- (0.33,4.3);
	
	\draw (0,4.5) rectangle (1,5.5);
	\draw[->,thick,-{Latex[round,length=2mm,width=2mm]}] (0.33,4.7) -- (0.33,5.3);
	
	\draw (2,0) rectangle (3,1); 
	\draw[->,thick,-{Latex[round,length=2mm,width=2mm]}] (2.33,0.2) -- (2.33,0.8); \draw[->,thick,-{Latex[round,length=2mm,width=2mm]}] (2.67,0.8) -- (2.67,0.2);
	\draw (2,1) rectangle (3,2); 
	\draw[->,thick,-{Latex[round,length=2mm,width=2mm]}] (2.33,1.2) -- (2.33,1.8); \draw[->,thick,-{Latex[round,length=2mm,width=2mm]}] (2.67,1.8) -- (2.67,1.2);
	\draw (2,2) rectangle (3,3);
	\draw[->,thick,-{Latex[round,length=2mm,width=2mm]}] (2.33,2.2) -- (2.33,2.8); \draw[->,thick,-{Latex[round,length=2mm,width=2mm]}] (2.67,2.8) -- (2.67,2.2);
	\draw (2,3.5) rectangle (3,4.5);
	\draw[->,thick,-{Latex[round,length=2mm,width=2mm]}] (2.33,3.7) -- (2.33,4.3);
	\draw (2,4.5) rectangle (3,5.5);
	\draw[->,thick,-{Latex[round,length=2mm,width=2mm]}] (2.33,4.7) -- (2.33,5.3);
	
	\node [shift={(0.5,-0.5)}] {Ni 1};
	\node [shift={(2.5,-0.5)}] {Ni 2};
	\node [shift={(1.5,6)}] {FM};
	\node [shift={(-0.5,1.5)}] {$t_{2g}$};
	\node [shift={(-0.5,4.5)}] {$e_g$};
	
	\draw[->,black!60!green,thick,-{Latex}] (0.7,2.7) .. controls (2.7,4.75) and (2.7,4.25) .. (0.7,2.6);
	\node[text=black!60!green] at (1.5,4) {1};
	
	
	\draw (4.5,0) rectangle (5.5,1); 
	\draw[->,thick,-{Latex[round,length=2mm,width=2mm]}] (4.83,0.2) -- (4.83,0.8); \draw[->,thick,-{Latex[round,length=2mm,width=2mm]}] (5.17,0.8) -- (5.17,0.2);
	\draw (4.5,1) rectangle (5.5,2); 
	\draw[->,thick,-{Latex[round,length=2mm,width=2mm]}] (4.83,1.2) -- (4.83,1.8); \draw[->,thick,-{Latex[round,length=2mm,width=2mm]}] (5.17,1.8) -- (5.17,1.2);
	\draw (4.5,2) rectangle (5.5,3);
	\draw[->,thick,-{Latex[round,length=2mm,width=2mm]}] (4.83,2.2) -- (4.83,2.8); \draw[->,thick,-{Latex[round,length=2mm,width=2mm]}] (5.17,2.8) -- (5.17,2.2);
	\draw (4.5,3.5) rectangle (5.5,4.5);
	\draw[->,thick,-{Latex[round,length=2mm,width=2mm]}] (4.83,3.7) -- (4.83,4.3);
	\draw (4.5,4.5) rectangle (5.5,5.5);
	\draw[->,thick,-{Latex[round,length=2mm,width=2mm]}] (4.83,4.7) -- (4.83,5.3);
	
	\draw (6.5,0) rectangle (7.5,1); 
	\draw[->,thick,-{Latex[round,length=2mm,width=2mm]}] (6.83,0.2) -- (6.83,0.8); \draw[->,thick,-{Latex[round,length=2mm,width=2mm]}] (7.17,0.8) -- (7.17,0.2);
	\draw (6.5,1) rectangle (7.5,2); 
	\draw[->,thick,-{Latex[round,length=2mm,width=2mm]}] (6.83,1.2) -- (6.83,1.8); \draw[->,thick,-{Latex[round,length=2mm,width=2mm]}] (7.17,1.8) -- (7.17,1.2);
	\draw (6.5,2) rectangle (7.5,3);
	\draw[->,thick,-{Latex[round,length=2mm,width=2mm]}] (6.83,2.2) -- (6.83,2.8); \draw[->,thick,-{Latex[round,length=2mm,width=2mm]}] (7.17,2.8) -- (7.17,2.2);
	\draw (6.5,3.5) rectangle (7.5,4.5);
	\draw[->,thick,-{Latex[round,length=2mm,width=2mm]}] (6.83,4.3) -- (6.83,3.7);
	\draw (6.5,4.5) rectangle (7.5,5.5);
	\draw[->,thick,-{Latex[round,length=2mm,width=2mm]}] (6.83,5.3) -- (6.83,4.7);

	\node [shift={(5.0,-0.5)}] {Ni 1};
	\node [shift={(7.0,-0.5)}] {Ni 2};
	\node [shift={(6.0,6)}] {AFM};
	\node [shift={(4.0,1.5)}] {$t_{2g}$};
	\node [shift={(4.0,4.5)}] {$e_g$};
	
	\draw[->,red,thick,-{Latex}] (4.95,5.03) .. controls (7.25,5.25) and (7.25,4.75) .. (4.95,4.97);
	\node[text=red] at (6.0,5.35) {3};
	
	\draw[->,blue,thick,-{Latex}] (4.9,2.8) .. controls (7.25,4.75) and (7.25,4.25) .. (4.9,2.7);
	\node[text=blue] at (6.0,4) {2};

\end{tikzpicture}

\caption{Schematic diagram of the major ferromagnetic and antiferromagnetic exchange processes in the $e_g$ and $t_{2g}$ band manifolds. The boxes indicate electron orbitals, and the loops indicate hopping pathways.}
\label{fig:process}
\end{figure}
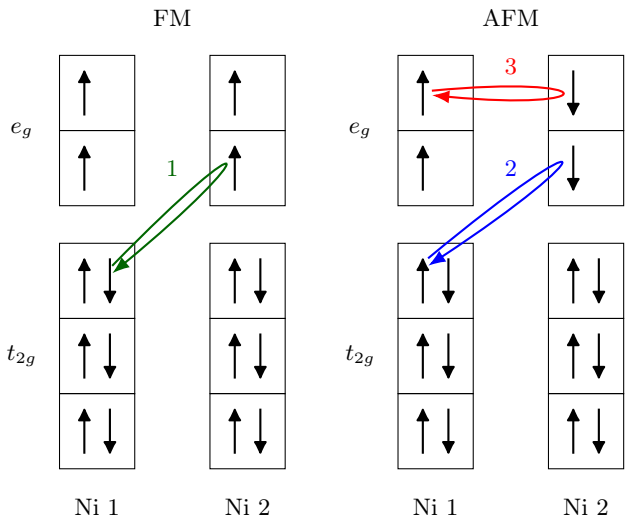

The theoretical prediction can be understood as follows. For a given pair of atoms, there are three major exchange processes to consider, one FM interaction and two AFM interactions as depicted in Figure~\ref{fig:process}. The FM (1) and AFM (2) processes involve $e_g$-$t_{2g}$ hopping, and AFM (3) process involves $e_g$-$e_g$ hopping. The difference between the former two processes is in the intermediate states which arise in the calculation of the second-order perturbation to the energy. The intermediate state of FM (1) maximizes the spin multiplicity (total spin quantum number) on one atom [in comparison to the AFM (2) process], giving a lower Hund's energy for that configuration, and thus a larger reduction in the energy. The FM (1) process tends to dominate for reasonable values of the interaction parameters. For the results in Table \ref{tab:FittedJ} we used $U = 3$~eV and $J_H = 0.5$~eV. See Appendix \ref{app:DFTmethods} for the full dependence on $U$ and $J_H$.

Thus, for the nearest neighbors where the strongest hopping is $e_g$-$t_{2g}$, the FM exchange process dominates and $J_1<0$. For the second-neighbors, the strongest hopping is again $e_g$-$t_{2g}$, and again FM dominates, but the maximum magnitude of these hopping integrals is less than half that of the nearest neighbors. So again $J_2<0$ (theoretically, and maybe experimentally to within uncertainty), but is less than one-quarter the magnitude of the nearest-neighbor exchange. Finally for the third-neighbors, the $e_g$-$e_g$ AFM process is by far the dominant exchange and $J_3>0$, with the largest magnitude of all the exchange constants. The third-neighbor $d_{z^2}$-$d_{xy}$ hopping integral is not negligible (Table~\ref{tab:hopping3N}), so the competition between the FM (1) and AFM (2) processes is still present, but this cannot overpower the $e_g$-$e_g$ AFM process for which the hopping integral is three times larger. However, this does indicate that a model including only the $e_g$-$e_g$ AFM process will overestimate the large $J_3$.

\section{Discussion}

At this point, we have a theoretical model for the observed spin waves and $Q=0$ excitations, as well as a first principles explanation for the strength of the exchange couplings. Thus we have answered our first question about the NiPS$_3$ exchange Hamiltonian.  Now we turn to the second question: given the proposed exotic Zhang-Rice behavior of NiPS$_3$, are there any features in the inelastic spectrum which can \textit{not} be accounted for by linear spin wave theory?

Although the LSWT calculation reproduces the inelastic spectrum well, the LSWT approach does not match the static ordered moment. Antiferromagnetic spin waves will, in general, reduce the size of the ground state ordered moment relative to its maximum value  \cite{ziman1972principles}, and substantial quantum entanglement can reduce the moment much further. Calculating the $T=0$ spin expectation value for the fitted spin wave Hamiltonian, we find $g \langle S \rangle = 1.73 \> \mu_B$ (assuming $g=2.00$) for Ni$^{2+}$. This is much larger than the experimental ordered moment 1.05~$\mu_B$ \cite{Wildes_2015}, which indicates that the real material NiPS$_3$ has substantially more quantum fluctuations than linear spin wave theory gives. 

In general, a strongly reduced static $T \rightarrow 0$ moment (formally defined by the ``one-tangle'' entanglement witness \cite{Scheie_2021_Witnessing}) indicates substantial quantum spin entanglement, showing that NiPS$_3$ is not merely a conventional antiferromagnet. In other words, the Ni$^{2+}$ magnetism cannot be described by linear spin wave theory alone, and therefore are subject to more exotic quantum effects. The missing spin components presumably reside within the excitation spectrum, potentially at $Q=0$ where neutrons cannot directly probe. 

The next item of comparison is details in the neutron spectrum. 
We compare the experimental data against resolution-convolved simulated scattering from best fit parameters in Table \ref{tab:FittedJ} using \textit{MCViNE}, a Monte Carlo ray tracing software to simulate time of flight resolution effects for the exact instrumental configuration and experimental bin widths \cite{Lin_2019,LIN201686} 
as shown in Fig. \ref{fig:ResConv} (see Appendix \ref{app:ResConv} for details).

\begin{figure}
	\centering
	\includegraphics[width=0.44\textwidth]{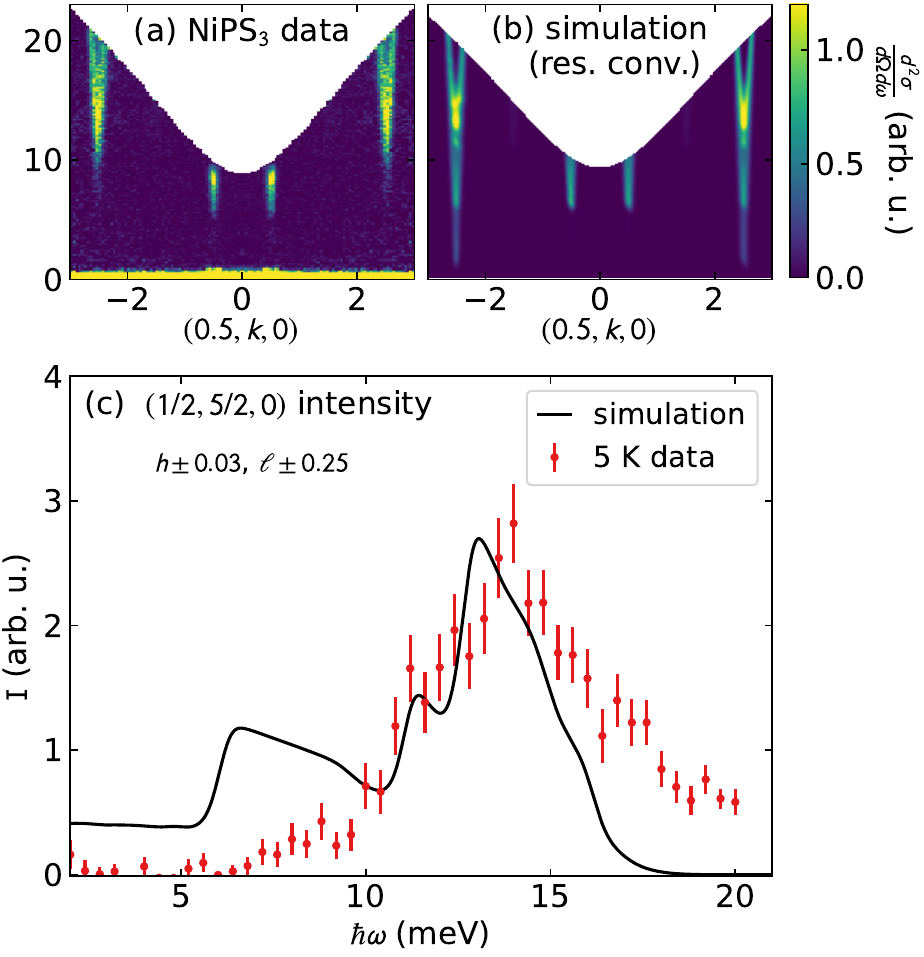}
	\caption{Comparison between simulated resolution broadened spin waves and experimental data at the bottom of the dispersion. Panel (a) shows the $E_i = 28$~meV data at 5~K, and panel (b) shows the exact same cut with the best fit Hamiltonian LSWT simulation with McVine simulated resolution convolution. The intensity maxima in experiment and theory are close in energy, but the  theoretical intensity extends much lower in energy. Panel (c) shows a constant $Q$ plot of experiment and simulation, showing that the experimental low energy tail of the dispersion is suppressed relative to the LSWT calculation.}
	\label{fig:ResConv}
\end{figure}

Although LSWT correctly captures the intensity near the finite-energy maximum, LSWT predicts a much larger low energy ``tail'' to the dispersion than is seen experimentally. 
In Appendix \ref{app:Su3} we show that quadrupolar $SU(3)$ dynamics (as will be present in $S=1$ Ni$^{2+}$ \cite{Zhang_2021_SUN}) explains one third of the reduced intensity, but nowhere near the dramatic reduction seen in experiment. The absent intensity must have a more exotic explanation. It is striking that the intensity is anomalously small at the lowest energies, near where the $\hbar \omega = 0$ static magnetism is also anomalously small. 
This means that quantum effects somehow seem to suppress the low-energy (long-time) magnetic response in NiPS$_3$. 

The combination of a reduced static moment and anomalously suppressed low energy intensity shows that LSWT fails to fully account for the low energy magnetism of NiPS$_3$. 
Further theoretical modeling is required to say for certain whether Zhang-Rice triplets account for the reduced moment, but we propose the observed resonances in Ref. \cite{Kang2020} as a potential explanation. A careful measure of the magnetic form factor could indicate whether a portion of the magnetic moment resides on the S sites in accord with Zhang-Rice triplet hypothesis. 
Be that as it may, these experimental observations beg for theoretical explanation: something very unusual is going on at the lowest energies. 
NiPS$_3$ has conventional magnons, that is not the end of the story.

\section{Conclusion}

We have measured the spin wave spectrum in NiPS$_3$ and modeled the spin waves, extracting a magnetic Hamiltonian with rigorously defined uncertainty. We have also used first principles calculations to model the magnetic exchange, and we find that DFT agrees very well with our experimentally determined exchange constants---in particular the anomalously large third neighbor exchange $J_3$. The microscopic mechanism for the dominant third neighbor exchange is elucidated by combining DFT with strong-coupling perturbation theory. 
Our fitted model is able to account for the finite energy maxima observed in neutron scattering, as well as the mode gaps observed in other experimental methods. 
The full profile we provide of long-wavelength ($Q=0$) magnetic excitations is essential knowledge for van der Waal magnets, because these modes most directly couple to optical and electronic excitations as relevant for spin-orbit entangled excitons and spintronic technology.  

Finally, we highlight a dramatically reduced static moment and suppressed low-energy intensity, which indicates that LSWT fails to fully explain NiPS$_3$ magnetism, 
especially in the low energy (long time) dynamics. 
This indicates an anomalous quantum state in NiPS$_3$, potentially driven by Zhang-Rice triplet pairing. 

(Note: In the final stages of this work, Ref. \cite{wildes2022magnetic} was published reporting similar measurements and a similar fitted spin exchange Hamiltonian to this study.)

%
%
\begin{acknowledgments} 

 This research used resources at the Spallation Neutron Source, a DOE Office of Science User Facility operated by the Oak Ridge National Laboratory. 
The work by A.S, J.V., C.L.S., and D.A.T. is supported by the Quantum Science Center (QSC), a National Quantum Information Science Research Center of the U.S. Department of Energy (DOE). 
S.O. is supported by the U.S. Department of Energy, Office of Science, Basic Energy Sciences, Materials Sciences and Engineering Division. Part of this research (T.B.) was conducted at the Center for Nanophase Materials Sciences, which is a DOE Office of Science User Facility. 
 Work on the resolution calculations was funded by the Laboratory Directors' Research and Development Fund of ORNL. 
 The resolution calculations also used resources of the Compute and Data Environment for Science (CADES) at the Oak Ridge National Laboratory, which is supported by the Office of Science of the U.S. Department of Energy under Contract No. DE-AC05-00OR22725.
 The work at SNU was supported by the Leading Researcher Program of Korea's National Research Foundation (Grant No. 2020R1A3B2079375). H.Z. gratefully acknowledges the support of the U.S. Department of Energy through the LANL/LDRD Program and the Center for Non-Linear Studies. 
 The authors acknowledge helpful discussions with Christian Batista. 
\end{acknowledgments} 

\appendix

\section{Sample preparation and experimental details}\label{app:ExpDetails}

Single-crystal NiPS3 was grown by a standard chemical vapor transport method. Pure Ni (>99.99\%), P (>99.99\%), and S (>99.998\%) powders were mixed in a molar ratio of 1:1:3 inside an Ar-filled glove box. We added an additional 5\% sulfur to the mixture for vapor transport. We analyzed the chemical composition of the resultant single crystals using energy-dispersive X-ray spectroscopy (Bruker QUANTAX 70), which confirmed a correct stoichiometry. We further characterized its magnetic property using a commercial SQUID magnetometer (MPMS-XL5, Quantum Design), the result of which is consistent with previous studies \cite{Wildes_2015,kim2019suppression}.

\begin{figure}
	\centering
	\includegraphics[width=0.40\textwidth]{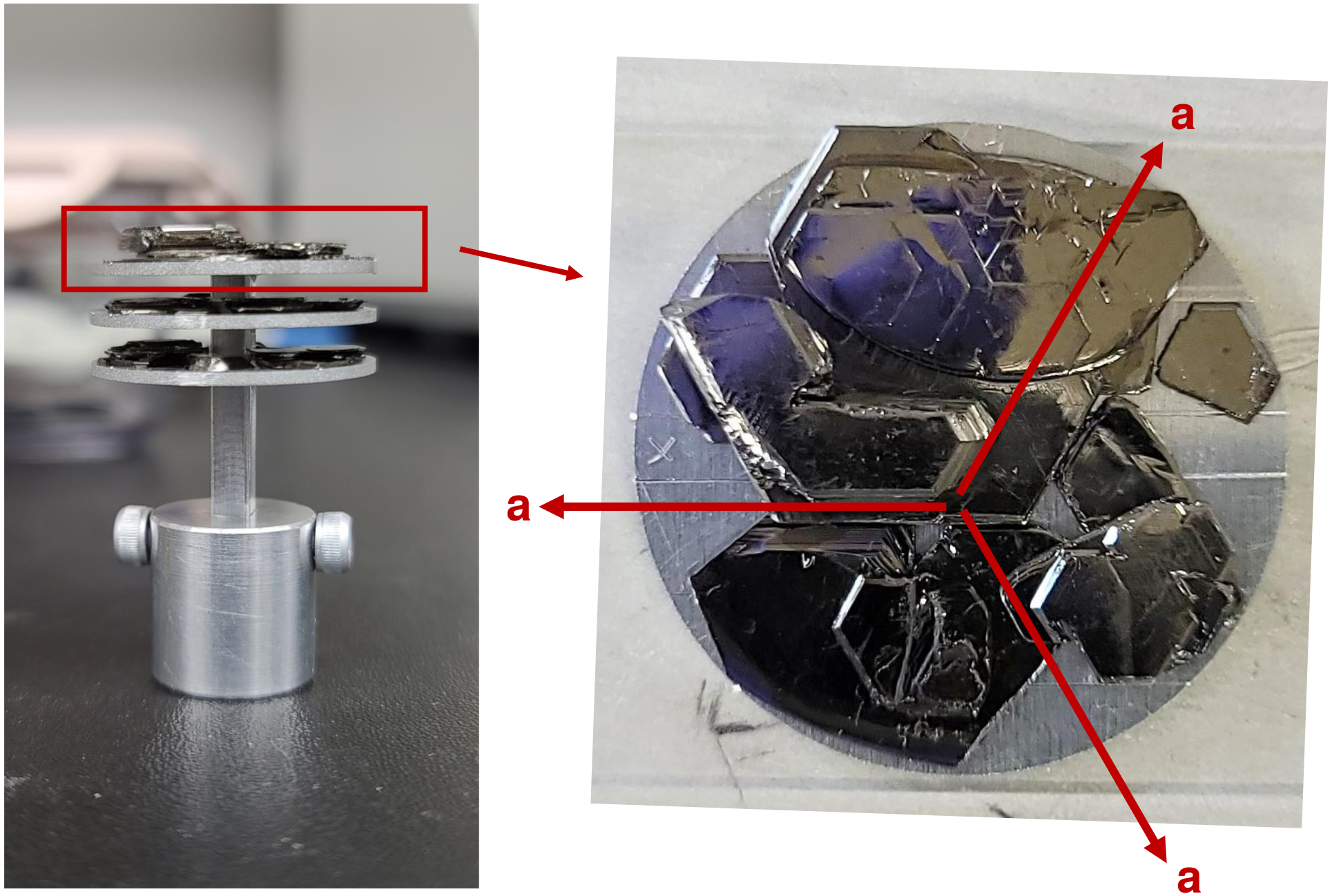}
	\caption{NiPS$_3$ sample mount, shown from the side and from the top. Because of the near three-fold rotation symmetry, the co-aligned crystals are aligned with the $a$ axis in one of three directions, as shown on the right.}
	\label{fig:samplemount}
\end{figure}

The sample for the SEQUOIA experiment was several coaligned NiPS$_3$ crystals totaling 2.41 g, glued to aluminum plates with CYTOP glue \cite{rule2018glue}. The sample mount in shown in Fig. \ref{fig:samplemount}. Because of the near three-fold rotation symmetry about ${\bf c}^{*}$ and the weak interplane van der Waals bonding, the sister compound FePS$_3$ has twinned domains separated by 120$^{\circ}$ rotation about ${\bf c}^{*}$ \cite{Chisato2016,Lancon_2016_twin}, and we expect the same situation with NiPS$_3$. Indeed, Xray Laue diffraction failed to distinguish the $[100]$ from the $[-1/2,1/2,0]$ or $[-1/2,-1/2,0]$ directions, which meant that the sample is a combination of orientations as shown in Fig. \ref{fig:samplemount}. 

\begin{figure}
	\centering
	\includegraphics[width=0.45\textwidth]{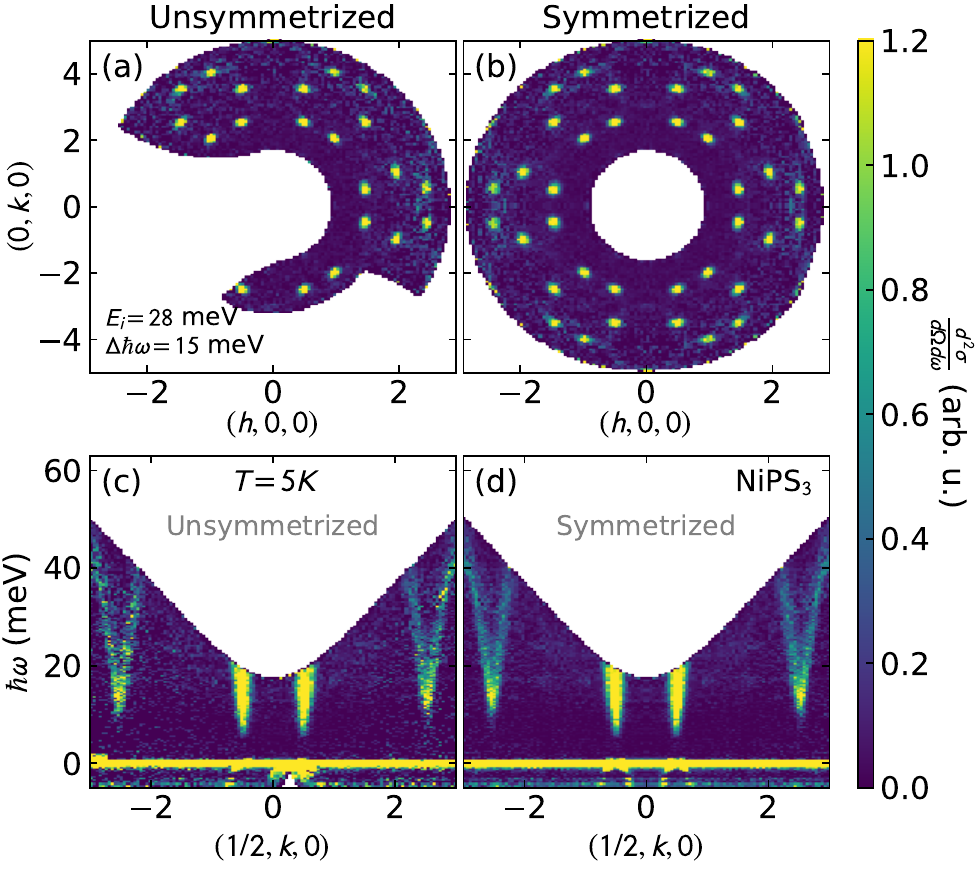}
	\caption{Symmetrization of NiPS$_3$ neutron data. The left column shows the unsymmetrized data, and the right shows the symmetrized data ($x,y,z;-x,y,z;x,-y,z;-x,-y,z$) for a constant energy $(hk0)$ slice, and a $(1/2,k,0)$ slice.}
	\label{fig:symmetrization}
\end{figure}

The instrument settings for the SEQUOIA neutron measurements are given in Table \ref{tab:SEQsettings}. 
For background, we made an identical sample holder with the same amount of CYTOP glued to it but with no crystals. This dummy sample was measured at the same energy and temperature configurations as the actual sample, and the measured scattering intensity was subtracted from the data. Plotted data were symmetrized with the following symmetry operations: $x,y,z;-x,y,z;x,-y,z;-x,-y,z$, see Fig. \ref{fig:symmetrization}.

\begin{table}
	\caption{SEQUOIA instrument parameters \cite{Granroth2010} for the NiPS$_3$ spin wave measurements at the various incident energies.  Fermi Chopper 2 (middle column) is the high-resolution chopper.}
	\begin{ruledtabular}
		\begin{tabular}{ccccc}
Nominal $E_i$ & Actual $E_i$ & Fermi & Fermi $\nu$ & $T_0$ $\nu$\\
(meV) & (meV)&  Chopper &(Hz) &(Hz) \\
\hline
100 & 103.4 & 2 & 540 &120 \\
60 & 62.1 & 2 & 420& 90 \\
28 & 28.9 & 2 & 300 & 60 \\
	\end{tabular}\end{ruledtabular}
	\label{tab:SEQsettings}
\end{table}

\begin{figure}
	\centering
	\includegraphics[width=0.47\textwidth]{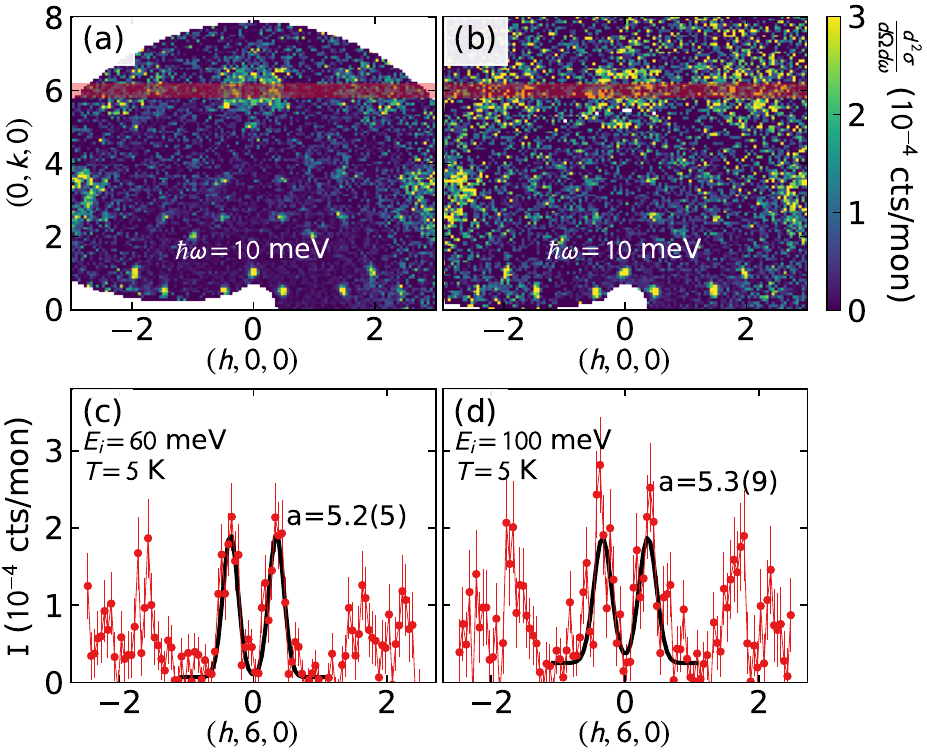}
	\caption{ Phonon fits for absolute unit conversions. Linear cuts through the $T=5$~K $(0,6,0)$ acoustic phonons at $\hbar \omega = 10$~meV scattering, the integrated width shown by the faint red lines in panels (a) and (b), at $E_i=60$~meV (left) and $E_i=100$~meV (right) were fitted to Gaussian curves to extract the area $a$ in panels (c) and (d). This was used to normalize the scattering intensity to absolute units.}
	\label{fig:AbsUnit}
\end{figure}

Data were normalized to absolute units by fitting the $(060)$ transverse acoustic phonon in accord with Ref. \cite{Xu_2013_AbsUnits} as shown in Fig. \ref{fig:AbsUnit}. Data were normalized per formula unit, equivalent to per Ni ion. 

\section{Linear spin wave fits} \label{app:SpinWaveFits}

In fitting the dispersions using linear spin wave theory, we extracted the mode energies at 267 unique $Q$ points, which are plotted in the supplemental information \cite{SuppMat}. The reduced $\chi^2$ of the $Q=0$ modes and the finite $Q$ spin wave modes were calculated separately and added, so that the number of points does not give undue weight to the neutron spectra. 

\begin{figure}
	\centering
	\includegraphics[width=0.48\textwidth]{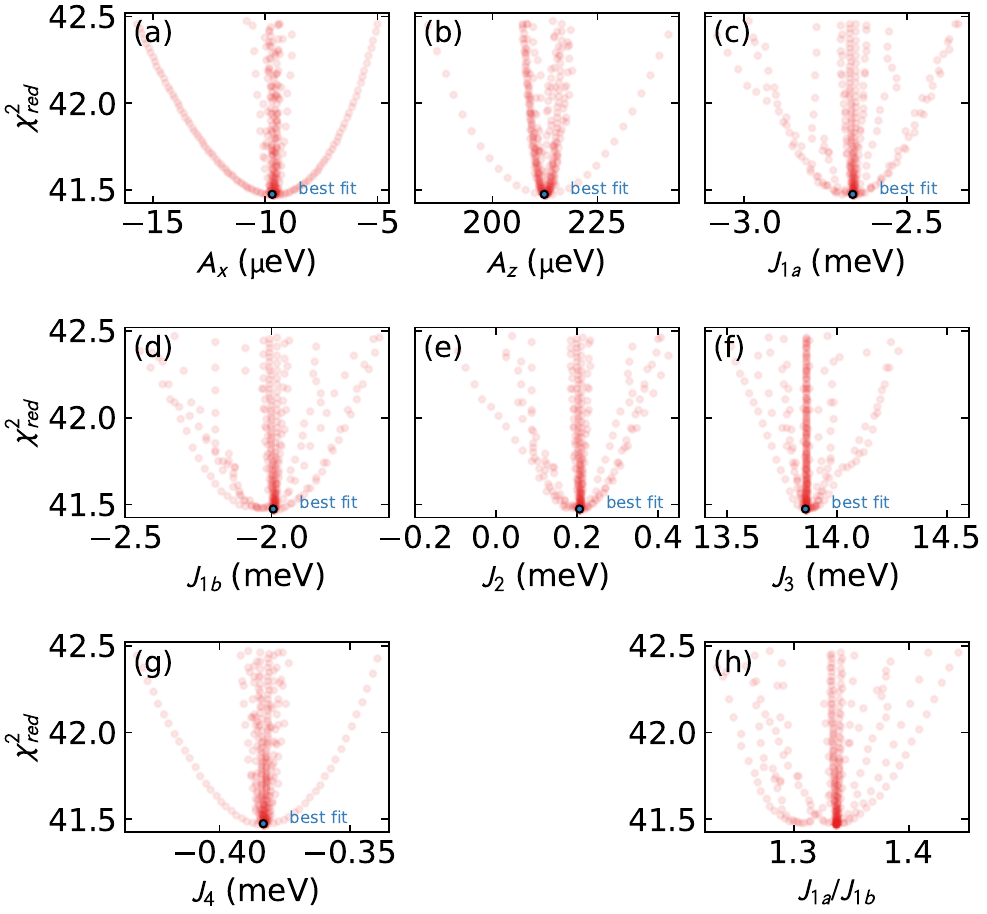}
	\caption{Range of solutions for NiPS$_3$ spin waves within $\Delta \chi^2 = 1$ of the best fit solution, using a method of fixing a parameter and allowing all others to fit freely. This was used to determine the one standard deviation uncertainty in Table \ref{tab:FittedJ}. Panel (h) shows the $\chi^2$ contour for $J_{1a}/J_{1b}$, showing that although the $J_{1a}$ and $J_{1b}$ single-value uncertainties overlap, they are unequal to within uncertainty.}
	\label{fig:uncertainty}
\end{figure}

We estimated uncertainty for the fitted exchange parameters by mapping out the reduced $\chi^2$ contour for one standard deviation uncertainty \cite{NumericalRecipes}. Following the method in Ref. \cite{Scheie_2022_GadoliniumPRB}, we fixed each parameter to a value slightly above or below its best fit value and varied the other parameters until an optimum fit was achieved. If this new best fit $\chi^2$ is within $\Delta \chi^2 = 1$ of the optimum $\chi^2$, we keep it as a valid solution and take another step away from the optimum. This is repeated until the best fit values are greater than  $\Delta \chi^2 = 1$, and are no longer within one standard deviation uncertainty of the optimum. Plots of valid solutions are shown in Fig. \ref{fig:uncertainty}. In this way, the extrema of the $\chi^2$ contour is mapped out along every fitted variable, and the extent is taken to be a measure of statistical uncertainty. 

\begin{figure}
	\centering
	\includegraphics[width=0.48\textwidth]{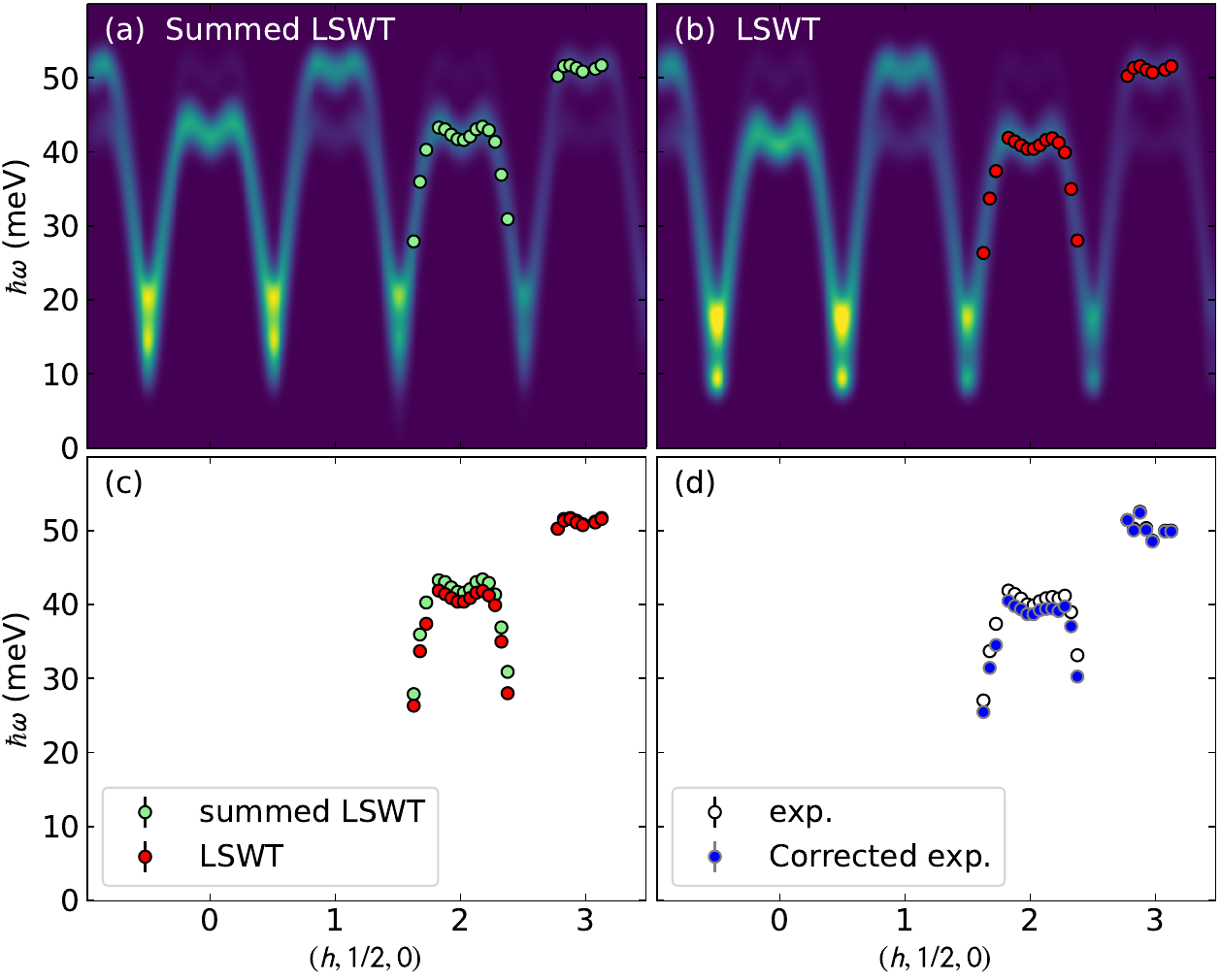}
	\caption{Effect of finite integration window on spin wave dispersion. Panel (a) shows the LSWT simulated scattering along $(h,1/2,0)$ summed over $-0.05<k<0.05$ reciprocal lattice units (RLU) and $-0.3 <\ell< 0.3$~RLU. Panel (b) shows the same data at exactly $(h,1/2,0)$, but with Gaussian broadening applied. The colored circles give the fitted mode energies at the same wavevectors that were extracted from experiment. Panel (c) shows the difference between the mode energies extracted from panels (a) and (b). Panel (d) shows the experimental extracted mode energies (white) and the corrected mode energies (blue) shifted by the offset determined in panel (c).}
	\label{fig:broadEffects}
\end{figure}

Fig. \ref{fig:broadEffects} shows the effect of finite width binning on the simulated LSWT data, and showing that this effect shifts the dispersions up in energy from their actual locations. In the final fits reported in the main text, the experimentally fitted spin wave modes were shifted downward in energy to account for this effect by calculating difference between the LSWT at infinitesimal $Q$ binning and at the actual experimental $Q$ binning. 

\subsection*{Effect of $J_3$}

\begin{figure*}
	\centering
	\includegraphics[width=0.97\textwidth]{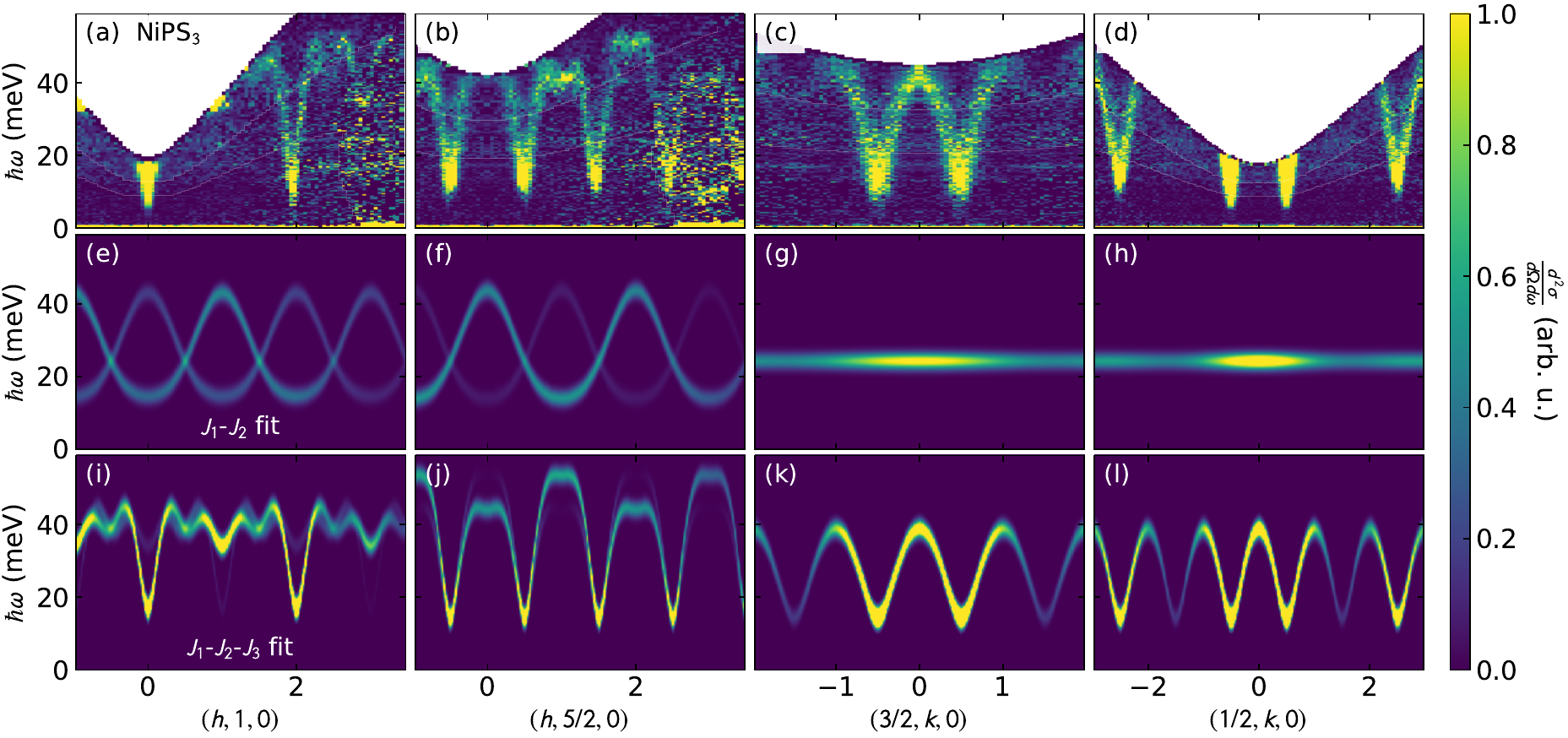}
	\caption{Measured and fitted NiPS$_3$ spin wave spectra showing the effect of $J_3$. The top row (a)-(e) shows the measured spin wave spectra. The second row (f)-(j) shows the fitted linear spin wave theory (LSWT) spectra using a $J_1$-$J_2$ model, and the bottom row shows the fit using a $J_1$-$J_2$-$J_3$  model. In none of the cuts do the $J_1$-$J_2$ model resemble the data, while the addition of $J_3$ makes the spin wave calculated modes match experiment much more closely.}
	\label{fig:spinWaveFitJ3}
\end{figure*}

 $J_3$ is by far the largest exchange interaction in the NiPS$_3$ Hamiltonian, and excluding $J_3$ from the fitted model worsens the a fit by an order of magnitude. To visually demonstrate the effect of $J_3$, we plot the best fit Hamiltonians both with and without $J_3$ in Figure \ref{fig:spinWaveFitJ3}. For certain cuts along $k$, nonzero $J_3$ is necessary to produce any dispersion at all, which in experiment is quite substantial.
Indeed, if we force $J_3$ to be zero and re-fit (including up to $J_5$), we find that the best fit  $\chi_{red}^2$ worsens by an order of magnitude ($\chi_{red}^2 = 41.5$ for $J_1$-$J_2$-$J_3$, $\chi_{red}^2 = 414.9$ with $J_3=0$). 
Thus the magnitude of $J_3$ is well-constrained by the experimental neutron scattering data.

\section{Resolution Convolution}\label{app:ResConv}

The instrumental resolution uses an incident beam profile calculated by Monte Carlo ray tracing in Mcstas with GPUs \cite{mcstas2020}.  This profile is then used to calculate point spread functions (PSF) on a discrete array across the slice by using the dgsres tool in \textit{McVINE} \cite{dgsresdoi}.  Next the PSFs are fit to provide parameters that allow interpolation of the resolution to any point in the slice \cite{Lin2022}.  The model slice to convolute was then calculated in \textit{SpinW} on a grid much finer than the resolution.  Finally, the interpolated functions were used to convolute the model slice with the instrumental resolution and produce the results. 

 For this specific slice the incident beam energy matching the measurement of $E_i=28.94$~meV was calculated \cite{ibeamdoi}.  The discrete array grid was along $k$ from -4 to 4 in steps of 0.4 and along $\hbar \omega$ from -5 to 26 in steps of 2 meV. The model slice was over the same bounds with 2036 $k$ bins and 1466 $\hbar \omega$ bins.

\section{Effects of $SU(3)$ dynammics}\label{app:Su3}

\begin{figure}
	\centering
	\includegraphics[width=0.48\textwidth]{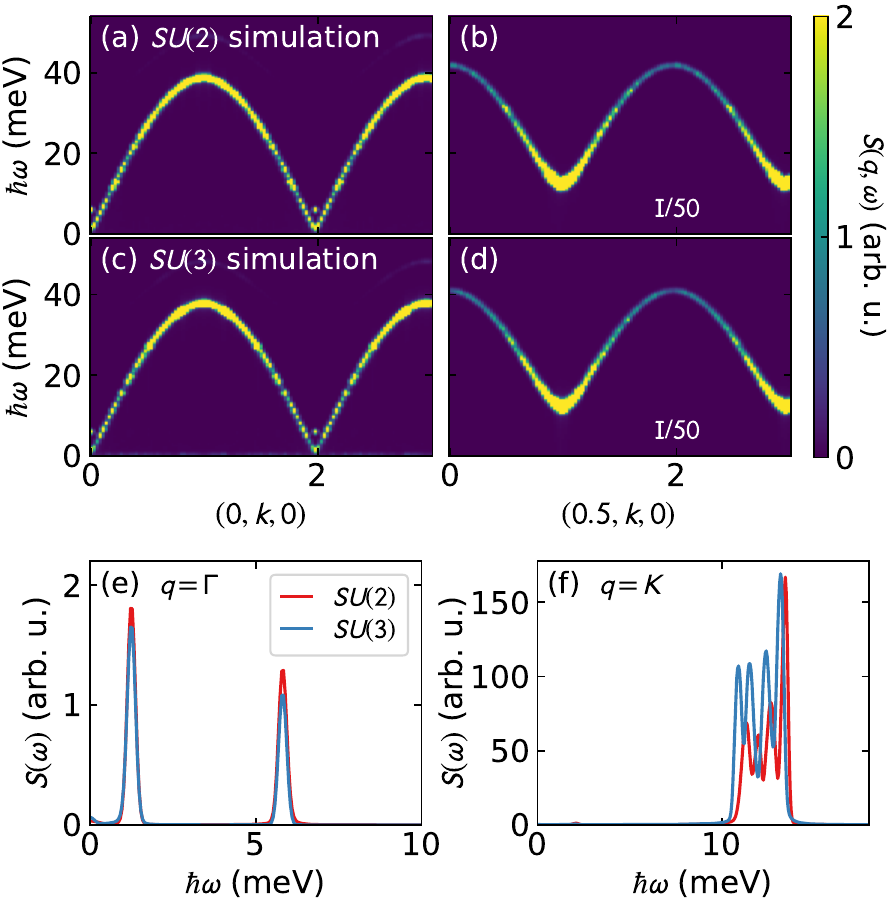}
	\caption{Calculated NiPS$_3$ dispersion for $SU(2)$ (classical dipolar) and $SU(3)$ ($S=1$) dynamics using Landau-Lifschitz dynamics as implemented with \textit{Su(n)ny} \cite{SUNNY}. The top panels show the calculated dispersions along two different cuts for  $SU(2)$ (a)-(b) and  $SU(3)$ (c)-(d). The differences are very minor, involving only a suppression of intensity at the lowest energy $q=\Gamma$ modes, and a slight enhancement at $q=K = (1/2,1/2,0)$. Thus quadripolar  $SU(3)$ dynamics does not explain the discrepancy between experiment and LSWT in Fig. \ref{fig:ResConv}.}
	\label{fig:Su3_comparison}
\end{figure}

As noted in the main text, a $S=1$ spin technically has $SU(3)$ symmetry. For weak anisotropies the $S=1$ spin can be treated as a dipole, but as anisotropy grows the higher order multipolar effects become more manifest, which allow a single-site spin singlet ($S=0$) state \cite{Zhang_2021_SUN}. 
To simulate the effects of this in NiPS$_3$, we calculated the inelastic neutron spectrum using the generalized spin wave package \textit{Su(n)ny} software suite \cite{SUNNY} using Landau-Lifshitz dynamics \cite{Dahlbom_2022} on a $75 \times 75 \times 4$ supercell at $T=5$~K using the fitted Hamiltonian in Table \ref{tab:FittedJ}. (In $SU(3)$ simulations the anisotropy was multiplied by two to keep the spin wave gaps at $\Gamma$ the same as $SU(2)$.) Both the $SU(3)$ and $SU(2)$ results are shown in Fig. \ref{fig:Su3_comparison}.  Note that the simulations in Fig. \ref{fig:Su3_comparison} do not include the effects of finite momentum space resolution.

The simulated  $SU(3)$ and $SU(2)$ spectra are nearly identical, involving only a weak 14\% suppression of low-energy intensity from higher order $SU(3)$ effects. If we normalize the low energy intensity relative to the 10-15 meV modes from $q=K$ (which practically speaking is what is done in Fig. \ref{fig:ResConv}), we find a suppression of 39\% in $SU(3)$ intensity relative to $SU(2)$. 
This is a mild reduction in intensity, but nowhere near as much as would be required to explain the absent intensity in Fig. \ref{fig:ResConv}. This means that the reduced low-energy intensity, alongside the reduced static magnetic moment, requires a more exotic explanation. 





\section{First principles calculations}\label{app:DFTmethods}
\begin{figure}
\includegraphics[width=0.48\textwidth]{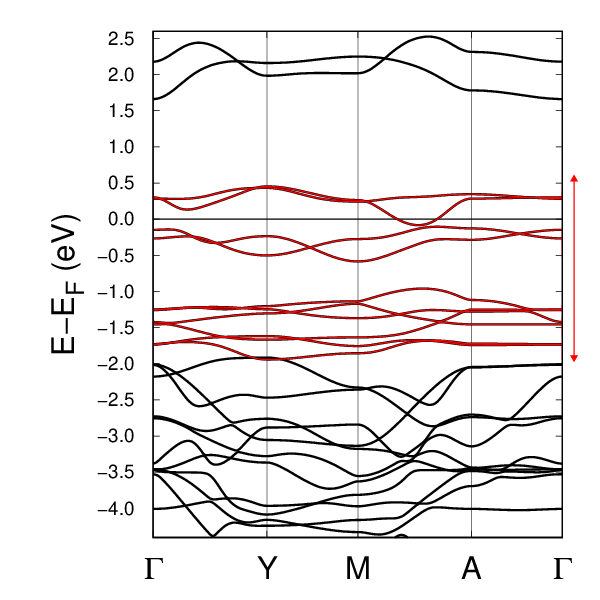}
\caption{Bulk band structure without spin-orbit coupling. The red bands are produced from the Wannier function Hamiltonian and the disentanglement window is depicted by the double-headed red arrow.}
\label{fig:bands}
\end{figure}

We perform density functional theory (DFT) calculations as implemented in VASP \cite{VASP1,VASP2}. The calculations are performed within the Perdew-Burke-Ernzerhof (PBE) generalized gradient approximation (GGA) \cite{GGA} for the exchange-correlation functional without spin-orbit coupling. We use projector augmented wave (PAW) pseudopotentials \cite{PAW1,PAW2} with an energy cutoff of 300 eV and an $11 \times 11 \times 9$ Monkhorst-Pack $k$-point mesh. We adopt the experimental lattice constants of Wildes \cite{Wildes_2015} for $C2/m$ bulk NiPS$_3$ and relax the atomic positions until component forces are less than 1 meV/\AA. We use Wannier90 \cite{W90,Marzari1997,Souza2001} to create a tight-binding Hamiltonian by projecting the band structure onto real Ni-$d$ orbitals. The maximal-localization step is not performed in order to maintain the symmetry of the Wannier functions close to their centers. The disentanglement window is shown by the double-headed arrow in Figure~\ref{fig:bands}(a) and the disentanglement convergence criterion is set to $10^{-13}$ \AA$^2$. The resulting Hamiltonian is ensured to be symmetrized by post-processing with WannSymm \cite{WannSymm}. 

The global Cartesian coordinate system was chosen such that the projection of the $z$-axis onto the Ni plane is perpendicular to the $Z_1$ bond \cite{Winter2016}. Explicitly, the primitive lattice row vectors for this choice of axes are
\begin{equation}
\begin{aligned}
\vec{a} &= (-2.3932999259, \hspace{0.75em}4.7774699422, -2.3841700163), \\
\vec{b} &= (-4.7774699422, \hspace{0.75em}2.3932999259,\hspace{0.75em} 2.3841700163), \\
\vec{c} &= (\hspace{0.75em}2.8698582203, \hspace{0.75em}2.8698582203, \hspace{0.75em}5.2691203152),
\end{aligned}
\end{equation}
in units of \AA.

Fig.~\ref{fig:bands} shows the excellent agreement between the electronic band structure calculated with DFT and from the Wannier tight-binding model for the Ni-$d$ orbitals. We accurately capture the $e_g$ bands near the Fermi level and the lower $t_{2g}$ bands.

\begin{figure}
\includegraphics[width=0.48\textwidth]{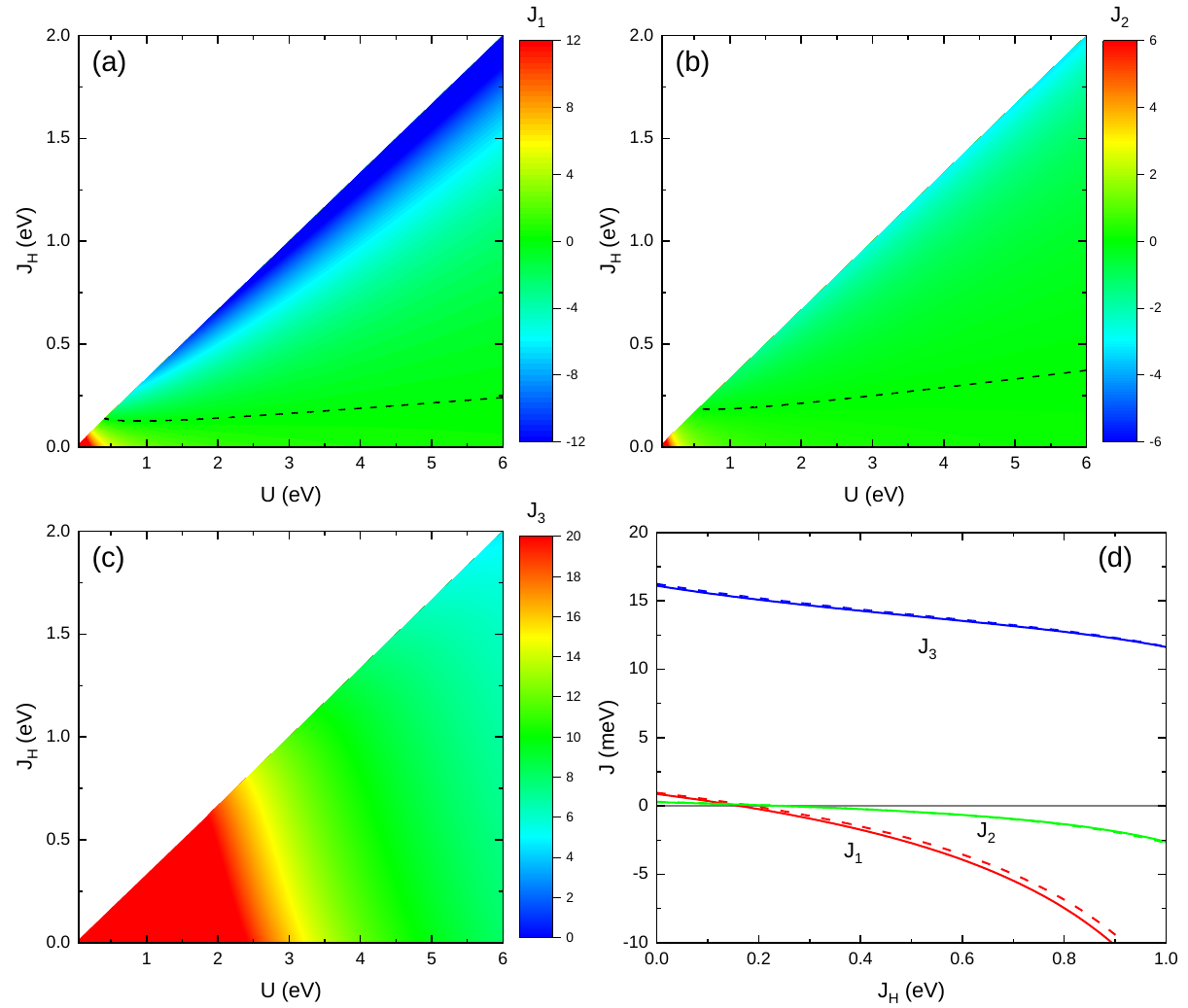}
\label{fig:satoshi}
\caption{Exchange constants (meV) as a function of $U$ and $J_H$ for (a) $J_1$ along the $X_1$ bond, 
(b) $J_2$ along the $X_2$ bond, and (c) $J_3$ along the $X_3$ bond. 
The dashed lines are the 0 meV isocurves. 
(d) For the specific case of $U=3~eV$, $J_1$ and $J_2$ are FM for all but the smallest $J_H$.
Solid lines are for $X_{1,2,3}$ bonds, and dashed lines are for $Z_{1,2,3}$ bonds.}
\end{figure}

To carry out the second-order perturbation calculation, the single-particle Hamiltonian as parameterized by a Wannier tight-binding model is
supplemented by a local Coulomb interaction Hamiltonian given by 
\begin{eqnarray}
H_U \! &=& \! U\sum_\alpha d^\dag_{\alpha \uparrow} d_{\alpha \uparrow} d^\dag_{\alpha \downarrow} d_{\alpha \downarrow}
+U' \sum_{\alpha \ne \beta} d^\dag_{\alpha \uparrow} d_{\alpha \uparrow} d^\dag_{\beta \downarrow} d_{\beta \downarrow} \nonumber \\
&&\!\! +(U'-J_H) \!\! \sum_{\alpha > \beta, \sigma} \!\! d^\dag_{\alpha \sigma} d_{\alpha \sigma} d^\dag_{\beta \sigma} d_{\beta \sigma} \nonumber \\
&&\!\! + J_H \sum_{\alpha \ne \beta} \Bigl( 
d^\dag_{\alpha \uparrow} d_{\beta \uparrow} d^\dag_{\beta \downarrow}d_{\alpha \downarrow}
+ d^\dag_{\alpha \uparrow} d_{\beta \uparrow} d^\dag_{\alpha \downarrow} d_{\beta \downarrow} \Bigr), 
\end{eqnarray}
where $\alpha$ and $\beta$ label $(yz,zx,xy,z^2,x^2-y^2)$ at Ni $d$ shell, 
and $d_{\alpha \sigma}^{(\dag)}$ is the annihilation (creation) operator of an electron at orbital $\alpha$ with spin $\sigma$. 
$U$ and $U'$ are the intraorbital Coulomb interaction and the interorbital Coulomb interaction, respectively, 
and $J_H$ represents the interorbital exchange interaction, i.e., the Hund coupling, (fourth term) and the interorbital pair hopping (fifth term). 
Between three parameters, we assume $U'=U-2J_H$ \cite{Kanamori_1963}. 
%
Because the energy scale of $U$ is order of eV, while that of the off-diagonal and anisotropic terms in the crystal field is smaller than 0.1~eV, except for the level difference between $e_g$ and $t_{2g}$ multiplets, known as $10Dq$. 
Thus, we only consider $10Dq$ by averaging $e_g$ and $t_{2g}$ levels
\begin{eqnarray}
H_{CF} = 10Dq  \sum_{\alpha \in e_g, \sigma} d^\dag_{\alpha \sigma} d_{\alpha \sigma}, 
\end{eqnarray}
with the average $t_{2g}$ level set to zero. 
The relativistic spin-orbit coupling is not included for simplicity. 

By diagonalizing $H_{CF}+H_U$ with $d^8$ configurations for Ni$^{2+}$ ion, we obtain the high-spin $e_g^2 t_{2g}^6$ ground state. 
From a pair of such high-spin $e_g^2 t_{2g}^6$ states, we proceed to 
carry out second-order perturbation calculations with respect to intersite electron hopping between Ni sites. 
Here, we consider two magnetic sites with ferromagnetic spin alignment ($E_{FM}$) and antiferromagnetic alignment ($E_{AFM}$) 
and compute the second-order correction to the ground state energy. 
During this process, all excited states for $d^7$ and $d^9$ configurations are included by diagonalizing $H_{CF}+H_U$. 
Finally, considering the full spin rotational symmetry, an exchange constant at a given pair of Ni spins is given by $J=(E_{FM}-E_{AFM})/2$. 

\begin{figure}
\includegraphics[width=0.36\textwidth]{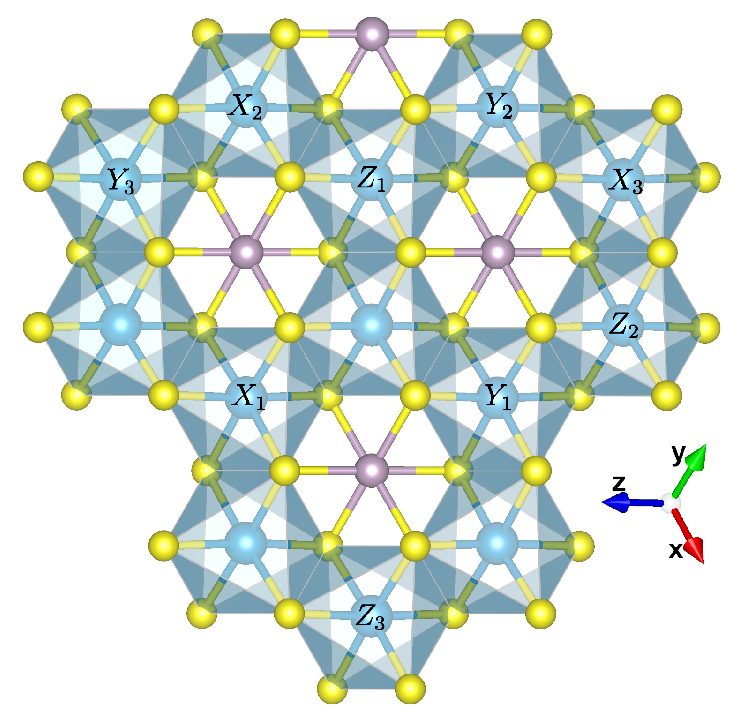}
\label{fig:hoplabels}
\caption{In-plane atoms to which the hopping integrals of Tables \ref{tab:hopping1N}, \ref{tab:hopping2N}, and \ref{tab:hopping3N} refer.}
\end{figure}

\break
\begin{table}
\caption{Local (on-site) hopping integrals (meV).}
\begin{center}
\begin{tabular}{c c c c c c}
\hline\hline
& \multicolumn{5}{c}{Local} \\
\cline{2-6}
& $d_{z^2}$ & $d_{x^2-y^2}$ & $d_{xz}$ & $d_{yz}$ & $d_{xy}$ \\
\hline
$d_{z^2}$ & 0.00 & 0.00 & -2.22 & -2.22 & 4.60 \\
$d_{x^2-y^2}$ & 0.00 & -17.22 & -3.40 & 3.40 & 0.00 \\
$d_{xz}$ & -2.22 & -3.40 & -1339.76 & 53.81 & 46.98 \\
$d_{yz}$ & -2.22 & 3.40 & 53.81 & -1339.76 & 46.98 \\
$d_{xy}$ & 4.60 & 0.00 & 46.98 & 46.98 & -1330.35 \\
\hline
\hline
\end{tabular}
\end{center}
\label{tab:hopping0N}
\end{table}

\begin{table*}
\caption{Nearest-neighbor hopping integrals (meV).}
\begin{center}
\begin{tabular}{c c c c c c c c c c c c c}
\hline\hline
& \multicolumn{5}{c}{$Z_1$ bond} & ~~&  \multicolumn{5}{c}{$X_1$ bond} \\
\cline{2-6} \cline{8-12}
& $d_{z^2}$ & $d_{x^2-y^2}$ & $d_{xz}$ & $d_{yz}$ & $d_{xy}$ & ~~  & $d_{z^2}$ & $d_{x^2-y^2}$ & $d_{xz}$ & $d_{yz}$ & $d_{xy}$\\
\hline
$d_{z^2}$ & -52.94 & 0.00 & -1.97 & -1.97 & 176.01 &~~ &-21.43 & 17.50 & 66.07 & -85.97 & -61.37 \\
$d_{x^2-y^2}$ & 0.00 & -10.72 & 67.39 & -67.39 &0.00 & ~~ & 17.50 & -39.73 & 34.68 & 157.39 & -36.97 \\
$d_{xz}$ & -1.97 & 67.39 & 45.49 & 13.28 & 34.00 &~~ &66.07 & 34.68 & 42.59 & 36.31 & 19.48 \\
$d_{yz}$ & -1.97 & -67.39 & 13.28 & 45.49 & 34.00 &~~ &-85.97 & 157.39 & 36.31 & -174.72 & 36.19 \\
$d_{xy}$ & 176.01 & 0.00 & 34.00 & 34.00 & -178.25 &~~ & -61.37 & -36.97 & 19.48 & 36.19 & 43.91\\
\hline
\hline
\end{tabular}
\end{center}
\label{tab:hopping1N}
\end{table*}

\begin{table*}
\caption{Second-neighbor hopping integrals (meV).}
\begin{center}
\begin{tabular}{c c c c c c c c c c c c c}
\hline\hline
& \multicolumn{5}{c}{$Z_2$ bond} & ~~&  \multicolumn{5}{c}{$X_2$ bond} \\
\cline{2-6} \cline{8-12}
& $d_{z^2}$ & $d_{x^2-y^2}$ & $d_{xz}$ & $d_{yz}$ & $d_{xy}$ & ~~  & $d_{z^2}$ & $d_{x^2-y^2}$ & $d_{xz}$ & $d_{yz}$ & $d_{xy}$\\
\hline
$d_{z^2}$ & 29.54 & 3.25 & -43.37 & 4.04 & 67.06 &~~ & 3.58 & -12.56 & 27.18 & -51.42 & -3.02 \\
$d_{x^2-y^2}$ & -3.25 & -5.17 & 26.69 & -31.79 & 19.83 & ~~ & -17.29 & 19.47 & 15.87 & 46.52 & -48.79 \\
$d_{xz}$ & 4.04& 31.79 & 5.15 & -44.68 & 9.16 &~~ & 46.00 & -22.06 & 5.20 & 11.64 & 20.57 \\
$d_{yz}$ & -43.37 & -26.69 & 18.10 & 5.15 & 11.79 &~~ & -17.11 & 70.16 & 7.20 & -2.98 & 12.50 \\
$d_{xy}$ & 67.06 & -19.83 & 11.79 & 9.16 & -3.50 &~~ & -28.69 & -14.71 & -42.31 & 8.83 & 4.80 \\
\hline
\hline
\end{tabular}
\end{center}
\label{tab:hopping2N}
\end{table*}

\begin{table*}
\caption{Third-neighbor hopping integrals (meV).}
\begin{center}
\begin{tabular}{c c c c c c c c c c c c c}
\hline\hline
& \multicolumn{5}{c}{$Z_3$ bond} & ~~&  \multicolumn{5}{c}{$X_3$ bond} \\
\cline{2-6} \cline{8-12}
& $d_{z^2}$ & $d_{x^2-y^2}$ & $d_{xz}$ & $d_{yz}$ & $d_{xy}$ & ~~  & $d_{z^2}$ & $d_{x^2-y^2}$ & $d_{xz}$ & $d_{yz}$ & $d_{xy}$\\
\hline
$d_{z^2}$ & -47.60 & 0.00 & 2.19 & 2.19 & 72.99 &~~ &152.31 & 111.35 & -6.38 & -35.50 & 5.70 \\
$d_{x^2-y^2}$ & 0.00 & 215.52 & -9.11 & -9.11 & 0.00 &~~ & 111.35 & 18.54 & -0.69 & 62.54 & 5.44 \\
$d_{xz}$ & 2.19 & -9.11 & 12.03 & -11.18 & -5.21 &~~ &-6.38 & -0.69 & 11.03 & -4.02 & -10.72 \\
$d_{yz}$ & 2.19 & -9.11 & -11.18 & 12.03 & -5.21 &~~ &-35.50 & 62.54 & -4.02 & 31.64 & -6.26 \\
$d_{xy}$ & 72.99 & 0.00 & -5.21& -5.21 & 32.74 &~~ & 5.70 & 5.44 & -10.72 & -6.26 & 11.37 \\
\hline
\hline
\end{tabular}
\end{center}
\label{tab:hopping3N}
\end{table*}

\clearpage

\end{document}